\documentclass[11pt]{article}
\usepackage{amsmath,amsthm,amssymb,color}
\usepackage{cite}
\usepackage[alwaysadjust]{paralist} 
\usepackage{lscape}

\newcommand{\ontop}[2]{\genfrac{}{}{0pt}{}{#1}{#2}}

\DeclareMathOperator{\tr}{tr}
\DeclareMathOperator{\Tr}{Tr}
\DeclareMathOperator{\Fix}{Fix}
\DeclareMathOperator{\Stab}{Stab}
\DeclareMathOperator{\swt}{swt}
\DeclareMathOperator{\wt}{wt}
\DeclareMathOperator{\w}{w}

\DeclareMathOperator{\puncture}{P}
\DeclareMathOperator{\pc}{P}
\DeclareMathOperator{\B}{BCH}
\DeclareMathOperator{\Hom}{Hom}
\DeclareMathOperator{\image}{image}

\newcommand{\C}{\mathbf{C}}
\newcommand{\F}{\mathbf{F}}
\newcommand{\Z}{\mathbf{Z}}

\newcommand{\RM}{{\mathcal{R}}}

\newcommand{\ket}[1]{|#1\rangle}
\newcommand{\bra}[1]{\langle #1|}
\newcommand{\idtwo}{\mathbf{1}_2}
\newcommand{\nix}[1]{}
\newcommand{\onemat}{\mathbf{1}}

\newcommand{\mbf}{\mathbf}
\newcommand{\ds}{\displaystyle}

\newcommand{\sdual}{{\bot_s}}
\newcommand{\adual}{{\bot_a}}
\newcommand{\hdual}{{\bot_h}}

\newcommand{\stacked}[2]{\begin{tabular}{c} #1\\[-1ex] \scriptsize #2\end{tabular}}

\def\srow#1 #2{(#1_1,\dots,#1_n|#2_1,\dots,#2_n)}
\def\<#1|#2>s{\langle #1|#2 \rangle_s}
\def\(#1|#2)a{\langle#1|#2\rangle_a}

\newtheorem{theorem}{Theorem}
\newtheorem{lemma}[theorem]{Lemma}
\newtheorem{proposition}[theorem]{Proposition}
\newtheorem{corollary}[theorem]{Corollary}

\theoremstyle{definition}

\newtheorem{example}[theorem]{Example}
\theoremstyle{remark}
\newtheorem{remark}[theorem]{Remark}

\begin{document}
\title{\Large\textbf{Nonbinary Stabilizer Codes over Finite Fields}}
\author{Avanti Ketkar\footnotemark, Andreas
Klappenecker\footnotemark, \setcounter{footnote}{1} Santosh Kumar\footnotemark[1],\\ Pradeep
Kiran Sarvepalli\\ Texas A\&M University, Department of Computer
Science,\\ College Station, TX 77843-3112} 

\renewcommand\thefootnote{\fnsymbol{footnote}}
\footnotetext[1]{Avanti Ketkar and Santosh Kumar are now with
Microsoft Corporation, Seattle.}
\footnotetext[2]{Contact author, e-mail: klappi at cs.tamu.edu} 
\date{}
\maketitle

\begin{abstract}
\noindent\footnotesize One formidable difficulty in quantum communication and
computation is to protect information-carrying quantum states against
undesired interactions with the environment. In past years, many
good quantum error-correcting codes had been derived as binary
stabilizer codes. Fault-tolerant quantum computation prompted the
study of nonbinary quantum codes, but the theory of such codes is not
as advanced as that of binary quantum codes.  This paper describes the
basic theory of stabilizer codes over finite fields.  The relation
between stabilizer codes and general quantum codes is clarified by
introducing a Galois theory for these objects.  A characterization of
nonbinary stabilizer codes over $\F_q$ in terms of classical codes
over $\F_{q^2}$ is provided that generalizes the well-known notion of
additive codes over $\F_4$ of the binary case. This paper derives
lower and upper bounds on the minimum distance of stabilizer codes,
gives several code constructions, and derives numerous families of
stabilizer codes, including quantum Hamming codes, quadratic residue
codes, quantum Melas codes, quantum BCH codes, and quantum character
codes.  The puncturing theory by Rains is generalized to additive
codes that are not necessarily pure. Bounds on the maximal length of
maximum distance separable stabilizer codes are given. A discussion of
open problems concludes this paper.
\end{abstract}

\begin{flushright}
{\small\textsl{This paper is dedicated to the memory of Professor
Thomas Beth}}
\end{flushright}

\section{Introduction}
\noindent
Reliable quantum information processing requires mechanisms to reduce
the effects of environmental and operational noise.  Fortunately, it
is possible to alleviate the detrimental effects of decoherence by
employing quantum error-correcting codes, so that one can engineer more
reliable quantum communication schemes and quantum computers.

The most widely studied class of quantum error-correcting codes are
binary stabilizer codes,
see~\cite{ashikhmin00a,ashikhmin00b,beth98,calderbank96,calderbank97,
cleve97,cleve97b,cohen99,danielsen05,ekert96,freedman01,gottesman02,
gottesman97,gottesman05,gottesman96b,grassl99,grassl99b,grassl00,
grassl01,kim02,kim03,kitaev97,martin04,rains99c,shor95,steane96,steane96b,
steane99,steane99b,thangaraj01,xiaoyan04} and, in particular, the seminal
works~\cite{calderbank98,gottesman96}.  An appealing aspect of binary
stabilizer codes is that there exist links to classical coding theory
which ease the construction of good codes.  More recently, some
results were generalized to the case of nonbinary stabilizer
codes~\cite{aharonov97,arvind03,ashikhmin01,bierbrauer00,chau97,chau97b,
feng02,feng02b,gottesman99,grassl03,grassl04,kim04,li04,matsumoto00,
rains99,roetteler04,schlingemann00,schlingemann02}, but the theory is
not nearly as complete as in the binary case.

We recall the basic principles of nonbinary stabilizer codes over
finite fields in the next section. In Section~3, we introduce a Galois
theory for quantum error-correcting codes.  The original theory
developed by Evariste Galois relates field extensions with
groups. Oystein Ore distilled the essence of this correspondence and
derived a significantly more general theory for pairs of
lattices~\cite{ore44}. We use this framework and set up a Galois
correspondence between quantum error-correcting codes and groups.
This theory shows how some properties of general quantum codes, such
as bounds on the minimum distance, can be deduced from results about
stabilizer codes.

In Section 4, we recall that stabilizer codes over a finite field
$\F_q$ correspond to additive codes over $\F_q$ that are
self-orthogonal with respect to a trace-symplectic
form~\cite{ashikhmin01}. We also establish the correspondence to
additive codes over $\F_{q^2}$ that are self-orthogonal with respect
to a trace-alternating form; remarkably, this basic construction had
been missing in the literature, in spite of the fact that it is a
generalization of the famous $\F_4$-codes~\cite{calderbank98}.

The MacWilliams relations for weight enumerators of stabilizer codes
are particularly easy to prove, as we show in Section 5. We then
derive in Section~6 upper and lower bounds on the minimum distance of
the best possible stabilizer codes. In Section 7, we recall basic
facts about cyclic stabilizer codes.

After laying the foundation in the first seven sections, we are able
to construct numerous code families in the subsequent sections.  In
Section 8, we derive quantum Hamming codes; in Section 9, quantum
quadratic residue codes; in Section 10, quantum Melas codes; and in
Section 11, quantum BCH codes. In the latter case, we show that
it is possible to extend quantum BCH codes. In Section 12, we
generalize the known results about puncturing pure linear stabilizer
codes to arbitrary additive codes, and we illustrate this theory by
puncturing quantum BCH codes.

We show in Section 13 that stabilizer codes over $\F_q$ attaining the
quantum Singleton bound cannot exceed a length of $q^2+1$, except in a
few sporadic cases, assuming that the classical MDS conjecture holds.
We give slightly weaker bounds for the length of MDS stabilizer
codes without such an assumption. In Section 14, we derive an
interesting class of quantum character codes. We give numerous code
constructions in Section 15, and conclude the paper with a discussion
of open questions.

We tried to keep the prerequisites to a minimum, so that readers from
the coding theory community as well as from the quantum computing
community can benefit. Apart from the basics of quantum computing, we
recommend \cite{calderbank98} and \cite{gottesman97} for background on
binary stabilizer codes, in addition to books on classical coding
theory, such as~\cite{huffman03} and~\cite{macwilliams77}. The general
theory of quantum codes is discussed in \cite{KnLa97}, and we assume
that the reader is familiar with the notion of a detectable error, as
introduced there. In general, we will omit proofs for results from our
companion papers~\cite{preprint0501126,klappenecker05p1}, but
otherwise we tried to make this paper reasonably self-contained.
\smallskip

\textit{Notations.}  We assume throughout this paper that $\F_q$
denotes a finite field of characteristic~$p$; in particular, $q$
always denotes a power of a prime $p$.  The trace function from
$\F_{q^m}$ to $\F_q$ is defined as $\tr_{q^m/q}(x)=\sum_{k=0}^{m-1}
x^{q^k}$; we may omit the subscripts if $\F_q$ is the prime field. If
$G$ is a group, then we denote by $Z(G)$ the center of this group. If
$S\subseteq G$, then we denote by $C_G(S)$ the centralizer of $S$ in
$G$.  We write $H\le G$ to express the fact that $H$ is a
subgroup of~$G$.  The trace~$\Tr(M)$ of a square matrix~$M$ is the sum of
the diagonal elements of~$M$.

\section{Stabilizer Codes}
Let $q$ a power of a prime $p$, and let $\C^q$ be a $q$-dimensional
complex vector space representing the states of a quantum mechanical
system. We denote by $\ket{x}$ the vectors of a distinguished
orthonormal basis of $\C^q$, where the labels $x$ range over the
elements of a finite field $\F_q$ with $q$ elements.  A quantum
error-correcting code $Q$ is a $K$-dimensional subspace of $\C^{q^n}=
\C^q\otimes \cdots \otimes \C^q$.

We need to select an appropriate error model so that we can measure
the performance of a code.  We simplify matters by choosing a basis
$\mathcal{E}_n$ of the vector space of complex $q^n\times q^n$
matrices to represent a discrete set of errors.  A stabilizer code is
defined as the joint eigenspace of a subset of~$\mathcal{E}_n$, so the
error operators play a crucial role.

\paragraph{Error Bases.}
Let $a$ and $b$ be elements of the finite field $\F_q$.  We define the
unitary operators $X(a)$ and $Z(b)$ on~$\C^q$ by
$$ X(a)\ket{x}=\ket{x+a},\qquad
Z(b)\ket{x}=\omega^{\tr(bx)}\ket{x},$$ where $\tr$ denotes the trace
operation from the extension field $\F_q$ to the prime field $\F_p$,
and $\omega=\exp(2\pi i/p)$ is a primitive $p$th root of unity.

We form the set $\mathcal{E}=\{X(a)Z(b)\,|\, a,b\in \F_q\}$ of error
operators. The set $\mathcal{E}$ has some interesting properties,
namely (a) it contains the identity matrix, (b)~the product of two
matrices in $\mathcal{E}$ is a scalar multiple of another element in
$\mathcal{E}$, and (c)~the trace $\Tr(A^\dagger B)=0$ for distinct
elements $A,B$ of $\mathcal{E}$. A finite set of $q^2$ unitary matrices
that satisfy the properties (a), (b), and (c) is called a \textsl{nice
error basis}, see~\cite{knill96a}.

The set $\mathcal{E}$ of error operators forms a basis of the set of
complex $q\times q$ matrices thanks to property (c).  We include a
proof that $\mathcal{E}$ is a nice error basis, because parts of our
argument will be of independent interest in the subsequent sections.

\begin{lemma}
The set $\mathcal{E}=\{X(a)Z(b)\,|\, a,b\in \F_q\}$ is a nice error
basis on\/~$\C^q$.
\end{lemma}
\begin{proof}
The matrix $X(0)Z(0)$ is the identity matrix, so property (a) holds.
We have $\omega^{\tr(ba)}X(a)Z(b)=Z(b)X(a)$, which implies that the
product of two error operators is given by
\begin{equation}\label{eq:multrule}
X(a)Z(b)\,X(a')Z(b')=\omega^{\tr(ba')}X(a+a')Z(b+b').
\end{equation}
This is a scalar multiple of an operator in $\mathcal{E}$,
hence property (b) holds.

Suppose that the error operators are of the form $A=X(a)Z(b)$ and
$B=X(a)Z(b')$ for some $a, b, b'\in \F_q$. Then
$$\Tr(A^\dagger B)= \Tr(Z(b'-b))=\sum_{x\in \F_q}
\omega^{\tr((b'-b)x)}.$$ The map $x\mapsto \omega^{\tr((b'-b)x)}$ is
an additive character of $\F_q$. The sum of all character values is 0
unless the character is trivial; thus, $\Tr(A^\dagger B)=0$ when
$b'\neq b$.

On the other hand, if $A=X(a)Z(b)$ and $B=X(a')Z(b')$ are two error
operators satisfying $a\neq a'$, then the diagonal
elements of the matrix $A^\dagger B=Z(-b)X(a'-a)Z(b')$ are 0, which
implies $\Tr(A^\dagger B)=0$.
Thus, whenever $A$ and $B$ are distinct element of $\mathcal{E}$, then
$\Tr(A^\dagger B)=0$, which proves~(c).
\end{proof}

\begin{example}
We give an explicit construction of a nice error basis with $q=4$
levels.  The finite field $\F_4$ consists of the elements $\F_4=\{0,
1, \alpha, \overline{\alpha}\}$. We denote the four standard basis
vectors of the complex vector space $\C^4$ by $\ket{0}, \ket{1},
\ket{\alpha},$ and $\ket{\overline{\alpha}}$.  Let $\idtwo$ denote the
$2\times 2$ identity matrix, $\sigma_x=\left(\begin{smallmatrix} 0&1\\
1&0
\end{smallmatrix}\right)$, and $\sigma_z=\left(\begin{smallmatrix}
1&\phantom{-}0\\ 0&-1\end{smallmatrix}\right)$. Then
\vspace*{-1ex}
$$
\begin{array}{c@{\,}c@{\,}lc@{\,}c@{\,}lc@{\,}c@{\,}lc@{\,}c@{\,}l}
X(0) &=& \idtwo\otimes \idtwo, &
X(1) &=& \idtwo\otimes \sigma_x, &
X(\alpha)&=&\sigma_x\otimes \idtwo, &
X(\overline{\alpha})&=& \sigma_x\otimes \sigma_x, \\
Z(0) &=& \idtwo\otimes \idtwo, &
Z(1) &=& \sigma_z\otimes \idtwo, &
Z(\alpha) &=& \sigma_z\otimes \sigma_z, &
Z(\overline{\alpha}) &=& \idtwo\otimes \sigma_z.
\end{array}
$$
We see that this nice error basis is obtained by tensoring the Pauli
basis, a nice error basis on $\C^2$. The next lemma shows that this is
a general design principle for nice error bases.
\end{example}

\begin{lemma}
If $\mathcal{E}_1$ and $\mathcal{E}_2$ are nice error bases, then
$$\mathcal{E}=\{ E_1\otimes E_2\,|\, E_1\in \mathcal{E}_1, E_2\in
\mathcal{E}_2\}$$ is a
nice error basis as well.
\end{lemma}
\noindent The proof of this simple observation follows directly from
the definitions.

Let $\mathbf{a}=(a_1,\dots, a_n)\in \F_q^n$. We write $ X(\mathbf{a})
= X(a_1)\otimes\, \cdots \,\otimes X(a_n)$ and $Z(\mathbf{a}) =
Z(a_1)\otimes\, \cdots \,\otimes Z(a_n)$ for the tensor products of
$n$ error operators.  Our aim was to provide an error model that
conveniently represents errors acting locally on one quantum
system. Using the new notations, we can easily formulate this model.

\begin{corollary}\label{th:nice}
The set
$\mathcal{E}_n= \{
X(\mathbf{a})Z(\mathbf{b})\,|\, \mathbf{a}, \mathbf{b}\in \F_q^n\}$ is
a nice error basis on the complex vector space~$\C^{q^n}$.
\end{corollary}

\textit{Remark.}  Several authors have used an error basis that is
equivalent to our definition of $\mathcal{E}_n$,
see~\cite{ashikhmin01,feng02b,kim04,matsumoto00}. We have defined the
operator $Z(b)$ in a slightly different way, so that the properties
relevant for the design of stabilizer codes become more
transparent. In particular, we can avoid an intermediate step that
requires tensoring $p\times p$--matrices, and that allows us to obtain
the trace-symplectic form directly, see Lemma~\ref{th:commute}.

\paragraph{Stabilizer Codes.}
Let $G_n$ denote the group generated by the matrices of the nice error
basis~$\mathcal{E}_n$.   It follows from
equation~(\ref{eq:multrule}) that
$$ G_n = \{ \omega^{c}X(\mathbf{a})Z(\mathbf{b})\,|\, \mathbf{a, b}
\in \F_q^n, c\in \F_p\}.$$ Note that $G_n$ is a finite group of order
$pq^{2n}$.  We call $G_n$ the \textsl{error group} associated with the nice
error basis $\mathcal{E}_n$.

A \textsl{stabilizer code} $Q$ is a non-zero subspace of $\C^{q^n}$
that satisfies
\begin{equation}\label{eq:stab}
Q = \bigcap_{E \in S} \{ v \in \C^{q^n} \mid Ev=v\}
\end{equation}
for some subgroup $S$ of $G_n$. In other words, $Q$ is the joint eigenspace to
the eigenvalue $1$ of a subgroup $S$ of the error group~$G_n$.
\smallskip

\textit{Remark.}  A crucial property of a stabilizer code is
that it contains \textsl{all} joint eigenvectors of~$S$ with
eigenvalue 1, as equation~(\ref{eq:stab}) indicates.  If the code is
smaller and does not exhaust all joint eigenvectors of $S$ with
eigenvalue 1, then it is not a stabilizer code for $S$.
\smallbreak

\paragraph{Minimum Distance.}
The error correction and detection capabilities of a quantum
error-correcting code $Q$ are the most crucial aspects of the code.
Recall that a quantum code~$Q$ is able to detect an error $E$ in the
unitary group $U(q^n)$ if and only if the condition $\langle c_1 | E|
c_2\rangle=\lambda_E\langle c_1 |c_2\rangle$ holds for all $c_1,
c_2\in Q$, see~\cite{KnLa97}.

It turns out that a stabilizer code~$Q$ with stabilizer $S$ can detect
all errors in $G_n$ that are scalar multiples of elements in $S$ or
that do not commute with some element of $S$, see
Lemma~\ref{th:detectable}. In particular, an error in $G_n$ that is
not detectable has to commute with all elements of the stabilizer.
Commuting elements in $G_n$ are characterized as follows:

\begin{lemma}\label{th:commute}
Two elements $E=\omega^cX(\mathbf{a})Z(\mathbf{b})$ and
$E'=\omega^{c'}X(\mathbf{a'})Z(\mathbf{b'})$ of the error group $G_n$
satisfy the relation
$$
EE' = \omega^{\tr(\mathbf{b\cdot a'-b'\cdot a})} E'E.
$$
In particular, the elements $E$ and $E'$ commute if and only if
the trace symplectic form
$\tr(\mathbf{b\cdot a'-b'\cdot a})$ vanishes.
\end{lemma}
\begin{proof}
It follows from equation (\ref{eq:multrule}) that
$EE'=\omega^{\tr(\mathbf{b\cdot a'})}X(\mathbf{a+a'})Z(\mathbf{b+b'})$
and $E'E=\omega^{\tr(\mathbf{b'\cdot a})}
X(\mathbf{a+a'})Z(\mathbf{b+b'})$. Therefore, multiplying $E'E$ with
the scalar $\omega^{\tr(\mathbf{b\cdot a'-b'\cdot a})}$ yields $EE'$,
as claimed.
\end{proof}

We define the \textsl{symplectic weight} $\swt$ of a vector $(\mbf a|\mbf b)$
in $\F_q^{2n}$ as
$$\swt((\mbf a|\mbf b)) = | \{\, k\, |\, (a_k,b_k)\neq (0,0)\}|.$$ The
weight $\w(E)$ of an element $E=\omega^c X(\mbf{a})Z(\mbf{b})$ in the
error group~$G_n$ is defined to be the number of nonidentity tensor
components, $\w(E)=\swt((\mbf a|\mbf b))$. In particular, the weight
of a scalar multiple of the identity matrix is by definition zero.

A quantum code $Q$ has \textsl{minimum distance} $d$\/ if and only if
it can detect all errors in $G_n$ of weight less than~$d$, but cannot
detect some error of weight~$d$.  We say that $Q$ is an $((n,K,d))_q$
code if and only if $Q$ is a $K$-dimensional subspace of $\C^{q^n}$
that has minimum distance~$d$.  An $((n,q^k,d))_q$ code is also called
an $[[n,k,d]]_q$ code. We remark that some authors are more
restrictive and use the bracket notation just to stabilizer codes.

We say that a quantum code $Q$ is \textsl{pure to} $t$\/ if and only
if its stabilizer group $S$ does not contain non-scalar matrices of
weight less than $t$.  A quantum code is called pure if and only if it
is pure to its minimum distance.  As in~\cite{calderbank98}, we
will always assume that an $[[n,0,d]]_q$ code has to be pure.  
\medskip

\textit{Remark.}  (a) If a quantum error-correcting code can detect a set
$\mathcal{D}$ of errors, then it can detect all errors in the linear
span of $\mathcal{D}$.  (b) A code of minimum distance~$d$ can correct all
errors of weight $t=\lfloor (d-1)/2\rfloor$ or less.

\section{Galois Connection}
We want to clarify the relation between stabilizer codes and more
general quantum codes before we proceed further.  Let us denote by
$\mathcal{Q}$ the set of all subspaces of $\C^{q^n}$. The set
$\mathcal{Q}$ is partially ordered by the inclusion relation.  Any two
elements of $\mathcal{Q}$ have a least upper bound and a greatest
lower bound with respect to the inclusion relation, namely
$$ \sup\{Q,Q'\} = Q+Q'\quad\mbox{and}\quad \inf\{Q,Q'\} = Q\cap Q'.$$
Therefore, $\mathcal{Q}$ is a complete (order) lattice. An element of this
lattice is a quantum error-correcting code or is equal to the vector
space $\{0\}$.

Let~$\mathcal{G}$ denote the lattice of subgroups of the error
group~$G_n$. We will introduce two order-reversing maps
between~$\mathcal{G}$ and $\mathcal{Q}$ that establish a Galois
connection. We will see that stabilizer codes are distinguished
elements of~$\mathcal{Q}$ that remain the same when mapped to the
lattice $\mathcal G$ and back.

Let us define a map $\Fix$ from the lattice $\mathcal{G}$ of subgroups
to the lattice $\mathcal{Q}$ of subspaces
that associates to a group $S$ its joint eigenspace with
eigenvalue~1,
\begin{equation}\label{eq:Fix}
\Fix(S) = \bigcap_{E\in S} \{ v\in \C^{q^n}\,|\, Ev=v\}.
\end{equation}
We define for the reverse direction a map $\Stab$ from the lattice
$\mathcal{Q}$ to the lattice $\mathcal{G}$ that associates to a
quantum code $Q$ its stabilizer group $\Stab(Q)$,
\begin{equation}\label{eq:Stab}
\Stab(Q) = \{ E\in G_n\,|\, Ev=v \text{ for all } v \in Q\}.
\end{equation}
We obtain four direct consequences of the definitions (\ref{eq:Fix})
and (\ref{eq:Stab}):
\begin{enumerate}
\item[\textbf{G1.}] If $Q_1\subseteq Q_2$ are subspaces of $\C^{q^n}$, then $\Stab(Q_2)\le \Stab(Q_1)$.
\item[\textbf{G2.}] If $S_1\le S_2$ are subgroups of $G_n$,
then $\Fix(S_2)\le \Fix(S_1)$.
\item[\textbf{G3.}] A subspace $Q$ of $\C^{q^n}$ satisfies
$Q\subseteq \Fix(\Stab(Q))$.
\item[\textbf{G4.}] A subgroup $S$ of $G_n$ satisfies
$S\le \Stab(\Fix(S))$.
\end{enumerate}
The first two properties establish that $\Fix$ and $\Stab$ are
order-reversing maps. The extension properties G3 and G4 establish
that $\Fix$ and $\Stab$ form a Galois connection,
see~\cite[page~56]{birkhoff61}. The general
theory of Galois connections establishes, among other results, that
$$ \Fix(S)=\Fix(\Stab(\Fix(S)))\quad\mbox{and}\quad
\Stab(Q)=\Stab(\Fix(\Stab(Q)))$$
holds for all $S$ in $\mathcal{G}$ and all $Q$ in $\mathcal{Q}$.

A subspace~$Q$ of the vector space~$\C^{q^n}$ satisfying G3 with
equality is called a \textsl{closed subspace}, and a subgroup $S$ of
the error group~$G_n$ satisfying G4 with equality is called a
\textsl{closed subgroup}.  We record the main result of abstract
Galois theory in the following proposition.

\begin{proposition}
The closed subspaces of the vector space~$\C^{q^n}$ form a complete
sublattice $\mathcal{Q}_c$ of the lattice~$\mathcal{Q}$. The closed
subgroups of\/ $G_n$ form a complete sublattice~$\mathcal{G}_c$ of the
lattice\/ $\mathcal G$ that is dual isomorphic to the
lattice~$\mathcal{Q}_c$.
\end{proposition}
\begin{proof}
This result holds for any Galois connection, see
Theorem~10 in the book by Birkhoff~\cite[page 56]{birkhoff61}.
\end{proof}

We need to characterize the closed subspaces and subgroups
to make this proposition useful. We begin with the closed subspaces
because this is easier.

\begin{lemma}
A closed subspace is a stabilizer code or is 0-dimensional.
\end{lemma}
\begin{proof}
By definition, a closed subspace $Q$ satisfies
$$ Q = \Fix(\Stab(Q)) =
\bigcap_{E\in \Stab(Q)} \{ v\in \C^{q^n}\,|\, Ev=v\},$$
hence is a stabilizer code or $\{0\}$.
\end{proof}

\begin{lemma}\label{th:stab}
If $Q$ is a nonzero subspace of $\C^{q^n}$, then its stabilizer
$S=\Stab(Q)$ is an abelian group satisfying $S\cap Z(G_n)=\{1\}$.
\end{lemma}
\begin{proof}
Suppose that $E$ and $E'$ are non-commuting elements of
$S=\Stab(Q)$. By Lemma~\ref{th:commute}, we have $EE'=\omega^k E'E$
for some $\omega^k\neq 1$.  A nonzero vector $v$ in~$Q$ would have to
satisfy $ v=EE'v=\omega^kE'Ev=\omega^kv,$ contradiction. Therefore,
$S$ is an abelian group. The stabilizer cannot contain any element
$\omega^k\onemat$, unless $k=0$, which proves the second assertion.
\end{proof}

\begin{lemma}\label{th:projection}
Suppose that $S$ is the stabilizer of a vector space $Q$.
An orthogonal projector onto the joint eigenspace $\Fix(S)$ is given by
$$ P= \frac{1}{|S|} \sum_{E\in S} E. $$
\end{lemma}
\begin{proof}
A vector $v$ in $\Fix(S)$ satisfies $Pv=v$, hence $\Fix(S)$ is
contained in the image of $P$. Conversely, note that $EP=P$ holds for
all $E$ in $S$, hence any vector in the image of $P$ is an eigenvector
with eigenvalue $1$ of all error operators $E$ in $S$. Therefore, $\Fix(S)=
\image P$. The operator $P$ is idempotent, because
$$ P^2 = \frac{1}{|S|} \sum_{E\in S} EP = \frac{1}{|S|} \sum_{E\in S} P = P$$
holds. The inverse $E^\dagger$ of $E$ is contained in the group $S$, hence $P^\dagger = P$. Therefore, $P$ is an orthogonal projector onto $\Fix(S)$.
\end{proof}

\textit{Remark.} If $S$ is a nonabelian subgroup of the group
$G_n$, then it necessarily contains the center $Z(G_n)$ of $G_n$; it
follows that $P$ is equal to the all-zero matrix. Note that the image
of $P$ has dimension $\Tr(P)=q^n/|S|$.

\begin{lemma}\label{th:closedsubgroup}
A subgroup $S$ of\/ $G_n$ is closed if and only if
$S$ is an abelian
subgroup that satisfies $S\cap Z(G_n)=\{1\}$ or if
$S$ is equal to $G_n$.
\end{lemma}
\begin{proof}
Suppose that $S$ is a closed subgroup of $G_n$. The vector space
$Q=\Fix(S)$ is, by definition, either a stabilizer code or a
0-dimensional vector space.  We have $\Stab(\{0\})=G_n$. Furthermore,
if $Q\neq \{0\}$, then $\Stab(Q)=S$ is an abelian group satisfying
$S\cap Z(G_n)=\{\onemat\}$, thanks to Lemma~\ref{th:stab}.

Conversely, suppose that $S$ is an abelian subgroup of $G_n$ such that
$S$ trivially intersects the center $Z(G_n)$.  Let
$S^*=\Stab(\Fix(S)).$ We have
$\Fix(S^*)=\Fix(\Stab(\Fix(S)))=\Fix(S),$ because this holds for any
pair of maps that form a Galois connection.
It follows from Lemma~\ref{th:projection} that
$$ q^n/|S^*| = \Tr\left(\frac{1}{|S^*|}\sum_{E\in S^*} E\right) =
\Tr\left(\frac{1}{|S|}\sum_{E\in S} E\right) = q^n/|S|.$$ 
Since $S\le S^*$, this shows that $S=S^*=\Stab(\Fix(S))$; hence, $S$ is
a closed subgroup of $G_n$. We note that $\Fix(G_n)=\{0\}$, so that
$G_n=\Stab(\Fix(G_n))$ is closed.
\end{proof}

The stabilizer codes are easier to study than arbitrary quantum codes,
as we will see in the subsequent sections. If we know the error
correction capabilities of stabilizer codes, then we get sometimes a
lower bound on the minimum distance of an arbitrary code by the
following simple observation: 
\smallskip

\noindent\textbf{Fact.} An arbitrary quantum code $Q$
is contained in the larger stabilizer code $Q^*=\Fix(\Stab(Q))$.  If
an error $E$ can be detected by $Q^*$, then it can be detected by $Q$
as well. Therefore, if the stabilizer code $Q^*$ has minimum distance
$d$, then the quantum code $Q$ has at least minimum distance $d$.

\section{Additive Codes}\label{sect:additive}
The previous section explored the relation between stabilizer codes
and other quantum codes. We show next how stabilizer codes are related
to classical codes (namely, additive codes over $\F_q$ or over
$\F_{q^2}$). The classical codes allow us to characterize the errors
in $G_n$ that are detectable by the stabilizer code.

If $S$ is a subgroup of $G_n$, then
$C_{G_n}(S)$ denotes centralizer of $S$ in $G_n$,
$$C_{G_n}(S)=\{ E\in G_n\,|\, EF=FE \text{ for all } F\in S\},$$ and
$SZ(G_n)$ denotes the group generated by $S$ and the center
$Z(G_n)$.  We first recall the following characterization of
detectable errors (see also~\cite{ashikhmin01}; the interested reader
can find a more general approach in~\cite{knill96b,klappenecker033}).

\begin{lemma}\label{th:detectable}
Suppose that $S \le G_n$ is the stabilizer group of a stabilizer
code~$Q$ of dimension $\dim Q>1$.  An error $E$ in $G_n$ is detectable by the
quantum code $Q$ if and only if either $E$ is an element of $SZ(G_n)$ or $E$
does not belong to the centralizer $C_{G_n}(S)$.
\end{lemma}
\begin{proof}
An element $E$ in $SZ(G_n)$ is a scalar multiple of a stabilizer;
thus, it acts by multiplication with a scalar $\lambda_E$ on $Q$.
It follows that $E$ is a detectable error.

Suppose now that $E$ is an error in $G_n$ that does not commute with some
element $F$ of the stabilizer~$S$; it follows that $EF= \lambda FE$ for some
complex number $\lambda \neq 1$, see Lemma~\ref{th:commute}.
All vectors $u$ and $v$ in $Q$ satisfy the condition
\begin{equation}\label{eq:noncommute}
\bra{u} E\ket{v} =\bra{u} EF\ket{v}=\lambda \bra{u} FE\ket{v}=
\lambda \bra{u}E\ket{v};
\end{equation}
hence, $\bra{u}E\ket{v}=0$. It follows that the error $E$ is detectable.

Finally, suppose that $E$ is an element of $C_{G_n}(S)\setminus
SZ(G_n)$.  Seeking a contradiction, we assume that $E$ is detectable;
this implies that there exists a complex scalar $\lambda_E$ such that
$Ev=\lambda_E v$ for all $v$ in $Q$.  The scalar $\lambda_E$ cannot be zero,
because $E$ commutes with the elements of $S$ so
$EP=PEP=\lambda_EP$ and clearly $EP\neq 0$.
Let $S^*$ denote the abelian
group generated by $\lambda_E^{-1} E$ and by the elements of $S$.  The
joint eigenspace of $S^*$ with eigenvalue 1 has dimension $q^n/|S^*| <
\dim Q=q^n/|S|$. This implies that not all vectors in $Q$ remain invariant under
$\lambda_E^{-1} E$, in contradiction to the detectability of $E$.
\end{proof}

\begin{corollary}
If a stabilizer code $Q$ has minimum distance $d$ and is pure to~$t$,
then all errors $E\in G_n$ with $1\le \wt(E)<\min\{t,d\}$ satisfy
$\langle u|E|v\rangle=0$ for all $u$ and $v$ in $Q$.
\end{corollary}
\begin{proof}
By assumption, the weight of $E$ is less than the minimum distance, so
the error is detectable. However, $E$ is not an element of $Z(G_n)S$,
since the code is pure to $t>\wt(E)$. Therefore, $E$ does not belong
to $C_{G_n}(S)$, and the claim follows from equation~(\ref{eq:noncommute}).
\end{proof}

\paragraph{Codes over $\F_q$.}
Lemma~\ref{th:detectable} characterizes the error detection capabilities of a
stabilizer code with stabilizer group $S$ in terms of the groups
$SZ(G_n)$ and $C_{G_n}(S)$. The phase information of an element in
$G_n$ is not relevant for questions concerning the detectability,
since an element $E$ of $G_n$ is detectable if and only if $\omega E$
is detectable. Thus, if we associate with an element
$\omega^cX(\mathbf{a})Z(\mathbf{b})$ of $G_n$ an element
$(\mathbf{a}|\mathbf{b})$ of $\F_q^{2n}$, then the group $SZ(G_n)$ is
mapped to the additive code
$$C=\{ (\mathbf{a}|\mathbf{b})\,|\,
\omega^cX(\mathbf{a})Z(\mathbf{b})\in SZ(G_n)\}= SZ(G_n)/Z(G_n).$$ To
describe the image of the centralizer, we need the notion of a
trace-symplectic form of two vectors $(\mathbf{a}|\mathbf{b})$ and
$(\mathbf{a'}|\mathbf{b'})$ in $\F_{q}^{2n}$,
$$ \< (\mathbf{a}|\mathbf{b})\, |\, (\mathbf{a'}|\mathbf{b'}) >s =
\tr_{q/p}(\mathbf{b}\cdot \mathbf{a}' - \mathbf{b}'\cdot
\mathbf{a}).$$ The centralizer $C_{G_n}(S)$ contains all elements of
$G_n$ that commute with each element of $S$; thus, by
Lemma~\ref{th:commute}, $C_{G_n}(S)$ is mapped onto
the trace-symplectic dual code $C^\sdual$ of the code $C$,
$$ C^\sdual
=\{ (\mathbf{a}|\mathbf{b})\,|\, \omega^cX(\mathbf{a})Z(\mathbf{b})\in C_{G_n}(S)\}.
$$
The connection between these classical codes and the stabilizer code
is made precise in the next theorem. This theorem is essentially
contained in~\cite{ashikhmin01} and generalizes the well-known
connection to symplectic codes~\cite{calderbank98,gottesman96} of the
binary case.

\begin{theorem}\label{th:stabilizer}
An $((n,K,d))_q$ stabilizer code exists if and only if there exists an
additive code $C \le \F_q^{2n}$ of size $|C|=q^n/K$ such that $C\le
C^\sdual$ and $\swt(C^{\sdual} \setminus C)=d$ if $K>1$ (and
$\swt(C^\sdual)=d$ if $K=1$).
\end{theorem}
\begin{proof}
Suppose that an $((n,K,d))_q$ stabilizer code $Q$ exists. This implies
that there exists a closed subgroup $S$ of $G_n$ of order $|S|=q^n/K$
such that $Q=\Fix(S)$.  The group $S$ is abelian and satisfies $S\cap
Z(G_n)=1$, by Lemma~\ref{th:closedsubgroup}.  The quotient $C\cong
SZ(G_n)/Z(G_n)$ is an additive subgroup of $\F_q^{2n}$ such that
$|C|=|S|=q^n/K$. We have $C^\sdual =
C_{G_n}(S)/Z(G_n)$ by Lemma~\ref{th:commute}.  Since $S$ is an abelian
group, $SZ(G_n)\le C_{G_n}(S)$, hence $C\le
C^\sdual$.
Recall that the weight of an element $\omega^c X(\mbf a)Z(\mbf b)$ in
$G_n$ is equal to $\swt(\mbf a|\mbf b)$. If $K=1$, then
$Q$ is a pure quantum code, thus $\wt(C_{G_n}(S))=\swt(C^\sdual)=d$.
If $K>1$, then the elements of
$C_{G_n}(S)\setminus SZ(G_n)$ have at least weight $d$ by
Lemma~\ref{th:detectable}, so that $\swt(C^\sdual \setminus C)=d$.

Conversely, suppose that $C$ is an additive subcode of
$\F_q^{2n}$ such that $|C|=q^n/K$, $C\le C^\sdual$, and
$\swt(C^\sdual\setminus C)=d$ if $K>1$ (and $\swt(C^\sdual)=d$ if $K=1$).
Let
$$ N = \{ \omega^cX(\mbf a)Z(\mbf b)\,|\, c \in \F_p \text{ and }
(\mbf a|\mbf b)\in C\}.$$ Notice that $N$ is an abelian normal
subgroup of $G_n$, because it is the pre-image of $C=N/Z(G_n)$.
Choose a character $\chi$ of $N$ such that
$\chi(\omega^c\onemat)=\omega^c$. Then
$$ P_N = \frac{1}{|N|}\sum_{E\in N} \chi(E^{-1}) E$$ is an orthogonal
projector onto a vector space $Q$, because $P_N$ is an idempotent in
the group ring $\C[G_n]$, see~\cite[Theorem~1]{klappenecker033}. We have
$$\dim Q = \Tr P_N = |Z(G_n)| q^n/|N| = q^n/|C| = K.$$ Each coset of
$N$ modulo $Z(G_n)$ contains exactly one matrix $E$ such that $Ev=v$
for all $v$ in $Q$. Set $S=\{E\in N\,|\, Ev=v \text{ for all } v \in
Q\}$. Then $S$ is an abelian subgroup of $G_n$ of order
$|S|=|C|=q^n/K$. We have $Q=\Fix(S)$, because $Q$ is clearly a
subspace of $\Fix(S)$, but $\dim Q=q^n/|S|=K$. An element $\omega^c
X(\mbf a)Z(\mbf b)$ in $C_{G_n}(S)\setminus SZ(G_n)$ cannot have
weight less than $d$, because this would imply that $(\mbf a|\mbf
b)\in C^\sdual\setminus C$ has weight less than $d$, which is
impossible. By the same token, if $K=1$, then all nonidentity elements
of the centralizer $C_{G_n}(S)$ must have weight $d$ or higher.
Therefore, $Q$ is an $((n,K,d))_q$ stabilizer code.
\end{proof}

\paragraph{Codes over $\F_{q^2}$.}
A drawback of the codes in the previous paragraph is that the
symplectic weight is somewhat unusual. In the binary case,
reference~\cite{calderbank98} provided a remedy by relating binary
stabilizer codes to additive codes over $\F_4$, allowing the use of
the familiar Hamming weight. Somewhat surprisingly, the corresponding
concept was not completely generalized to~$\F_{q^2}$, although
\cite{matsumoto00,kim04} and~\cite{rains99} paved the way to our
approach.  After circulating a first version of this manuscript,
Gottesman drew our attention to another interesting approach that was
initiated by Barnum, see~\cite{barnum00,barnum02}, where a sufficient
condition for the existence of stabilizer codes is established using a
symplectic form.
\smallskip

Let $(\beta,\beta^q)$ denote a normal basis of $\F_{q^2}$ over $\F_q$.
We define a trace-alternating form of two vectors $v$ and $w$ in
$\F_{q^2}^n$ by
\begin{equation}\label{eq:alternating}
 \(v|w)a =
\tr_{q/p}\left(\frac{v\cdot w^q - v^q\cdot w }{\beta^{2q}-\beta^2}\right).
\end{equation}
We note that the argument of the trace is invariant under the
Galois automorphism $x\mapsto x^q$, so it is indeed an element
of~$\F_q$, which shows that (\ref{eq:alternating}) is well-defined.

The trace-alternating form is bi-additive, that is,
$\(u+v|w)a = \(u|w)a+\(v|w)a $ and $\(u|v+w)a = \(u|v)a+\(u|w)a$ holds
for all $u,v,w\in \F_{q^2}^n$. It is $\F_p$-linear, but not
$\F_q$-linear unless $q=p$. And it is alternating in the sense that
$\(u|u)a=0$ holds for all $u\in \F_{q^2}^n$. We write $u \adual w$ if
and only if $\(u|w)a=0$ holds.

We define a bijective map $\phi$ that takes an element $(\mbf a|\mbf b)$ of the
vector space $\F_q^{2n}$ to a vector in $\F_{q^2}$ by setting
$\phi((\mbf a|\mbf b)) = \beta \mbf a + \beta^q \mbf b.$ The map
$\phi$ is isometric in the sense that the symplectic weight of
$(\mbf a|\mbf b)$ is equal to the Hamming weight of $\phi((\mbf a|\mbf b))$.

\begin{lemma}\label{th:isometry}
Suppose that $c$ and $d$ are two vector
of\/ $\F_q^{2n}$. Then
$$\< c\,|\,d>s=\(\phi(c)\,|\, \phi(d))a.$$ In particular, $c$ and $d$ are
orthogonal with respect to the trace-symplectic form if and
only if $\phi(c)$ and $\phi(d)$ are orthogonal with respect to the
trace-alternating form.
\end{lemma}
\begin{proof}
Let $c=(\mbf a|\mbf b)$ and $d=(\mbf a'|\mbf b')$. We calculate
$$
\begin{array}{lcl}
\phi(c)\cdot \phi(d)^q &=&
\beta^{q+1} \,\mbf a\cdot \mbf a' +
\beta^{2}   \,\mbf a\cdot \mbf b' +
\beta^{2q}  \,\mbf b\cdot \mbf a' +
\beta^{q+1} \,\mbf b\cdot \mbf b' \\[1ex]
\phi(c)^q\cdot \phi(d) &=&
\beta^{q+1} \, \mbf a\cdot \mbf a' +
\beta^{2q}  \, \mbf a\cdot \mbf b' +
\beta^2     \, \mbf b\cdot \mbf a' +
\beta^{q+1} \, \mbf b\cdot \mbf b'
\end{array}
$$
Therefore, the trace-alternating form of $\phi(c)$ and $\phi(d)$ is given by
$$ \(\phi(c)|\phi(d))a =
\tr_{q/p}\left(\frac{\phi(c)\cdot \phi(d)^q - \phi(c)^q\cdot \phi(d) }{\beta^{2q}-\beta^2}\right)=\tr_{q/p}(\mbf b \cdot \mbf a' - \mbf a \cdot \mbf b' ),
$$
which is precisely the trace-symplectic form $\< c\,|\,d>s$.
\end{proof}

\begin{theorem}\label{th:alternating}
An\/ $((n,K,d))_q$ stabilizer code exists if and only if there exists
an additive subcode $D$ of\/ $\F_{q^2}^{n}$ of cardinality
$|D|=q^n/K$ such that $D\le D^\adual$ and
$\wt(D^{\adual} \setminus D)=d$ if $K>1$ (and $\wt(D^\adual)=d$ if $K=1$).
\end{theorem}
\begin{proof}
Theorem~\ref{th:stabilizer} shows that an $((n,K,d))_q$ stabilizer
code exists if and only if there exists a code $C\le \F_q^{2n}$ with
$|C|=q^n/K$, $C\le C^\sdual$, and $\swt(C^\sdual\setminus C)=d$ if
$K>1$ (and $\swt(C^\sdual)=d$ if $K=1$). We
obtain the statement of the theorem by applying the isometry $\phi$.
\end{proof}

We obtain the following convenient condition for the existence of a
stabilizer code as a direct consequence of the previous theorem.
\begin{corollary}\label{co:alternating}
If there exists a classical\/ $[n,k]_{q^2}$ additive code $D\le
\F_{q^2}$ such that $D\le D^\adual$ and $d^\adual=\wt(D^\adual)$
then there exists an $[[n,n-2k,\geq d^\adual]]_{q}$ stabilizer 
code that is pure to~$d^\adual$.
\end{corollary}

\textit{Remark.} It is not necessary to use a normal basis in the
definition of the isometry $\phi$ and the trace-alternating
form. Alternatively, we could have used a polynomial basis $(1,\gamma)$
of $\F_q^2/\F_q$. In that case, one can define the isometry $\phi$ by
$\phi((\mathbf{a}|\mathbf{b}))=\mathbf{a}+\gamma \mathbf{b}$,
and a compatible trace-alternating form by
$$ \langle v\, | \, w\rangle_{a'} = \tr_{q/p}\left(\frac{
v \cdot w^q - v^q\cdot w
}{\gamma-\gamma^q}\right).$$
One can check that the statement of Lemma~\ref{th:isometry}
is satisfied for this choice as well.
Other variations on this theme are possible.

\paragraph{Classical codes.}
Self-orthogonal codes with respect to the trace-alter\-nating form are
not often studied in classical coding theory; more common are codes
which are self-orthogonal with respect to a euclidean or hermitian
inner product.  We relate these concepts of orthogonality in this
paragraph.

Consider the hermitian inner product $\mathbf{x}^q\cdot \mathbf{y}$ of
two vectors $\mathbf{x}$ and $\mathbf{y}$ in $\F_{q^2}^n$; we write
$\mathbf{x} \,\hdual\, \mathbf{y}$ if and only if $\mathbf{x}^q \cdot
\mathbf{y}=0$ holds.
\begin{lemma}\label{th:hermitian}
If two vectors $\mathbf{x}$ and $\mathbf{y}$ in $\F_{q^2}^n$ satisfy
$\mathbf{x} \, \hdual\, \mathbf{y}$, then they satisfy $\mathbf{x}
\,\adual\, \mathbf{y}$.  In particular, if\/ $D\le \F_{q^2}^n$, then
$D^\hdual \le D^\adual$.
\end{lemma}
\begin{proof}
It follows from $\mathbf{x}^q\cdot \mathbf{y}=0$ that
 $\mathbf{x}\cdot \mathbf{y}^q=0$ holds, whence
$$\(\mathbf{x}|\mathbf{y})a =
\tr_{q/p}\left(
\frac{\mathbf{x}\cdot \mathbf{y}^q - \mathbf{x}^q\cdot \mathbf{y}}{\beta^{2q}-\beta^2}\right)=0,$$
as claimed.
\end{proof}
Therefore, any self-orthogonal code with respect to the hermitian
inner product is self-orthogonal with respect to the trace-alternating
form. In general, the two dual space $D^\hdual$ and $D^\adual$ are not
the same. However, if $D$ happens to be $\F_{q^2}$-linear, then the
two dual spaces coincide.
\begin{lemma}\label{th:classical}
Suppose that $D\le \F_{q^2}^n$ is $\F_{q^2}$-linear, then $D^\hdual=D^\adual$.
\end{lemma}
\begin{proof}
Let $q=p^m$, $p$ prime. If $D$ is a $k$-dimensional subspace of
$\F_{q^2}^n$, then $D^\hdual$ is a $(n-k)$-dimensional subspace of
$\F_{q^2}^n$. We can also view~$D$ as a $2mk$-dimensional subspace of
$\F_p^{2mn}$, and $D^\adual$ as a $2m(n-k)$-dimensional subspace of
$\F_p^{2mn}$.  Since $D^\hdual \subseteq D^\adual$ and the
cardinalities of $D^\adual$ and $D^\hdual$ are the same, we can
conclude that $D^\adual =D^\hdual$.
\end{proof}

\begin{corollary}\label{co:classical}
If there exists an $\F_{q^2}$-linear $[n,k,d]_{q^2}$ code $B$ such
that $B^\hdual\le B$, then there exists an $[[n,2k-n,\geq d]]_{q}$
quantum code that is pure to~$d$.
\end{corollary}
\begin{proof}
The hermitian inner product is nondegenerate, so the hermitian dual of
the code $D:=B^\hdual$ is $B$.  The
$[n,n-k]_{q^2}$ code $D$ is $\F_{q^2}$-linear, so
$D^\hdual=D^\adual$ by Lemma~\ref{th:classical}, and the claim follows
from Corollary~\ref{co:alternating}.
\end{proof}

So it suffices to consider hermitian forms in the case of
$\F_{q^2}$-linear codes. We have to use the slightly more cumbersome
trace-alternating form in the case of additive codes that are not
linear over $\F_{q^2}$.

An elegant and surprisingly simple construction of quantum codes was
introduced in 1996 by Calderbank and Shor~\cite{calderbank96} and by
Steane~\cite{steane96}. The CSS code construction provides perhaps the
most direct link to classical coding theory.

\begin{lemma}[CSS Code Construction]\label{th:css}
Let $C_1$ and $C_2$ denote two classical linear codes with parameters
$[n,k_1,d_1]_q$ and $[n,k_2,d_2]_q$ such that $C_2^\perp\le C_1$.
Then there exists a $[[n,k_1+k_2-n,d]]_q$ stabilizer code with minimum
distance $d=\min\{ \wt(c) \mid c\in (C_1\setminus C_2^\perp)\cup
(C_2\setminus C_1^\perp)\}$ that is pure to $\min\{ d_1,d_2\}$.
\end{lemma}
\begin{proof}
Let $C=C_1^\perp\times C_2^\perp\le \F_q^{2n}$. If $(c_1\mid c_2)$ and
$(c_1'\mid c_2')$ are two elements of $C$, then we observe that
$$ \tr( c_2\cdot c_1' - c_2' \cdot c_1) = \tr(0-0)=0.$$ Therefore,
$C\le C^\sdual$. Furthermore, the trace-symplectic dual of $C$
contains $C_2\times C_1$, and a dimensionality argument shows that
$C^\sdual = C_2\times C_1$. Since the cartesian product
$C_1^\perp\times C_2^\perp$ has $q^{2n-(k_1+k_2)}$ elements, the
stabilizer code has dimension $q^{k_1+k_2-n}$ by
Theorem~\ref{th:stabilizer}. The claim about the minimum distance and purity of
the code is obvious from the construction.
\end{proof}

\begin{corollary}\label{th:css2}
If $C$ is a classical linear $[n,k,d]_q$ code containing its dual,
$C^\perp\le C$, then there exists an $[[n,2k-n,\geq d]]_q$ stabilizer code 
that is pure to~$d$.
\end{corollary}

\section{Weight Enumerators}
The Shor-Laflamme weight enumerators of an arbitrary $((n,K))_q$
quantum code~$Q$ with orthogonal projector $P$ are defined by the
polynomials
$$
\begin{array}{ll}
\ds \sum_{i=0}^n A_i^{\textsc{sl}} z^i, &\quad\text{with}\quad
\ds A_i^{\textsc{sl}}=\frac{1}{K^2}\sum_{\ontop{E\in G_n}{\wt(E)=i}} 
\Tr(E^\dagger P)\Tr(E P),
\end{array}
$$
and
$$
\begin{array}{ll}
\ds \sum_{i=0}^n B_i^{\textsc{sl}} z^i, &\quad\text{with}\quad
\ds B_i^{\textsc{sl}} = \frac{1}{K}\sum_{\ontop{E\in G_n}{\wt(E)=i}}
\Tr(E^\dagger PE P),\phantom{gggd}
\end{array}
$$ see~\cite{shor97} for the binary case.  The weights
$A_i^{\textsc{SL}}$ and $B _i^{\textsc{SL}}$ have a nice combinatorial
interpretation in the case of stabilizer codes.  Indeed, let $C\le
\F_q^{2n}$ denote the additive code associated with the stabilizer
code~$Q$.  Define the symplectic weights of $C$ and $C^\sdual$
respectively by
$$ A_i=|\{ c\in C\,|\, \swt(c)=i\}| \quad\text{and}\quad
B_i=|\{ c\in C^\sdual\,|\, \swt(c)=i\}|.$$
The next lemma belongs to the folklore of stabilizer codes. 

\begin{lemma}
The Shor-Laflamme weights of an $((n,K))_q$ stabilizer code $Q$ are
multiples of the symplectic weights of the associated additive
codes~$C$ and~$C^\sdual$; more precisely,
$$ A_i^\textsc{sl}=pA_i\quad \text{and}\quad B_i^\textsc{sl}=pB_i
\quad\text{for}\quad  0\le i\le n,$$
where $p$ is the characteristic of the field\/ $\F_q$.
\end{lemma}
\begin{proof}
Recall that
$$ P = \frac{1}{|S|} \sum_{E\in S} S$$ for the stabilizer group $S$ of
$Q$. The trace $\Tr(EP)$ is nonzero if and only if $E^\dagger$ is an
element of $SZ(G_n)$. If $E^\dagger\in SZ(G_n)$, then $\Tr(E^\dagger
P)\Tr(EP)=(q^n/|S|)^2=K^2$. Therefore, $A_{i}^\textsc{sl}$
counts the elements in $SZ(G_n)$ of weight $i$, so
$A_i^\textsc{sl}=|Z(G_n)|\times |\{c\in C\,|\, \swt(c)=i\}|=p A_i.$

If $E$ commutes with all elements in $S$, then $\Tr(E^\dagger PE
P)=\Tr(P^2)=\Tr(P)=K$. If $E$ does not commute with some element of
$S$, then $E$ is detectable; more precisely, the proof of
Lemma~\ref{th:detectable} shows that $PEP=0 P$, hence $\Tr(E^\dagger
PEP)=0$.  Therefore, $B_i^\textsc{sl}$ counts the elements in
$C_{G_n}(S)$ of weight $i$, hence $B_i^\textsc{sl} = |Z(G_n)| \times
|\{c\in C^\sdual\,|\, \swt(c)=i\}| = p A_i'.$
\end{proof}

Shor and Laflamme had been aware of the stabilizer case when they
introduced their weight enumerators, so the combinatorial
interpretation of the weights does not appear to be a coincidence.
Recall that the Shor-Laflamme enumerators of arbitrary quantum codes
are related by a MacWilliams identity, see~\cite{rains98,shor97}. For
stabilizer codes, we can directly relate the symplectic weight
enumerators of $C$ and $C^\sdual$,
$$ A(z) = \sum_{i=0}^n A_iz^i\quad\text{and}\quad B(z)=\sum_{i=0}^n
B_i z^i,$$ using a simple argument that is very much in the spirit of Jessie
MacWilliams' original proof for euclidean dual codes~\cite{macwilliams63}.

\begin{theorem}
Let $C$ be an additive subcode of $\F_q^{2n}$
with symplectic weight enumerator
$A(z)$. Then the symplectic weight enumerator of $C^\sdual$
is given by
$$ B(z) = \frac{(1+(q^2-1)z)^n}{|C|} A
\left(\frac{1-z}{1+(q^2-1)z)}\right). $$
\end{theorem}
\begin{proof}
Let $\chi$ be a nontrivial additive character of $\F_p$.  We define
for $b\in \F_q^{2n}$ a character $\chi_b$ of the additive group $C$ by
substituting the trace-symplectic form for the argument of
the character $\chi$, such that
$$ \chi_b(c) = \chi(\<c|b>s ). $$ The character $\chi_b$ is trivial if
and only if $b$ is an element of $C^{\sdual}$.
Therefore, we obtain from the orthogonality relations of characters that
$$ \sum_{c\in C} \chi_b(c) =
\left\{ \begin{array}{ll} |C| &
\text{ for } b\in C^{\sdual},\\
0 & \text{ otherwise.}
\end{array}\right.
$$
The following relation for polynomials is an immediate consequence
\begin{equation}\label{eq:mac}
\sum_{c\in C}\sum_{b\in \F_q^{2n}} \chi_b(c)z^{\swt(b)}  =
\sum_{b\in \F_q^{2n}} z^{\swt(b)} \sum_{c\in C} \chi_b(c) =
|C| B(z).
\end{equation}
The right hand side is a multiple of the weight enumerator of the code
$C^\sdual$. Let us have a closer look at the inner sum of the
left-hand side.  If we express the vector $c\in C$ in the form
$c=(c_1,\dots,c_n|d_1,\dots,d_n)$, and expand the character and its
trace-symplectic form, then we obtain
$$
\begin{array}{lc@{}l}
\ds\sum_{b\in \F_q^{2n}}  \chi_b(c)  z^{\swt(b)} &=& \ds\!\!
\sum_{(a_1,\dots,a_n|b_1,\dots,b_n)\in \F_q^{2n}}  \!\!\!\!\!\!
z^{\sum_{k=1}^n \swt(a_k|b_k)}
\chi\left(\sum_{k=1}^n \tr(d_ka_k-b_kc_k)\right)\\
&=& \ds\sum_{(a_1,\dots,a_n|b_1,\dots,b_n)\in \F_q^{2n}}
\prod_{k=1}^n z^{\swt(a_k|b_k)} \chi\left( \tr(d_ka_k-b_kc_k)\right) \\
&=&\; \ds\prod_{k=1}^n \sum_{(a_k|b_k)\in \F_q^2} z^{\swt(a_k|b_k)}
\chi\left( \tr(d_ka_k-b_kc_k)\right).
\end{array}
$$
Recall that $\chi$ is a nontrivial character of $\F_p$,
hence the map $(a_k|b_k)\mapsto \chi(\tr(d_ka_k-b_kc_k))$ is a
nontrivial character of $\F_q^2$ for all $(c_k|d_k)\neq (0|0)$.
Therefore, we can simplify the inner sum to
$$
\sum_{(a_k|b_k)\in \F_q^2} z^{\swt(a_k|b_k)}
\chi\left( \tr(d_ka_k-b_kc_k)\right) =
\left\{ \begin{array}{l@{\,}l}
1+(q^2-1)z & \text{ if } (c_k|d_k)=(0,0),\\
1-z & \text{ if } (c_k|d_k)\neq (0,0).
    \end{array}
\right.
$$
It follows that
$$
\sum_{b\in \F_q^{2n}} \chi_b(c)z^{\swt(b)}  =
(1-z)^{\swt(c)}(1+(q^2-1)z)^{n-\swt(c)}.$$
Substituting this expression into equation~(\ref{eq:mac}), we find that
$$
\begin{array}{lcl}
B(z) &=&\ds |C|^{-1}
\sum_{c\in C}\sum_{b\in \F_q^{2n}} \chi_b(c)z^{\swt(b)} \\
&=& \ds\frac{(1+(q^2-1)z)^n}{|C|}
\sum_{c\in C} \left(\frac{1-z}{1+(q^2-1)z}\right)^{\swt(c)}\\
&=& \ds\frac{(1+(q^2-1)z)^n}{|C|}\, A\!\left(\frac{1-z}{1+(q^2-1)z}\right),
\end{array}
$$
which proves the claim.
\end{proof}

The coefficient of $z^j$ in $(1+(q^2-1)z)^{n-x}(1-z)^x$ is given by
the Krawtchouk polynomial of degree $j$ in the variable $x$,
$$ K_j(x) =
\sum_{s=0}^j (-1)^s(q^2-1)^{j-s} {x \choose s}{n-x \choose j-s}.$$
\begin{corollary}\label{th:krawtchouk}
Keeping the notation of the previous theorem, we have
$$ B_j = \frac{1}{|C|}\sum_{x=0}^n K_j(x)A_x.$$
\end{corollary}
\begin{proof}
According to the previous theorem, we have
$$ \begin{array}{lcl}
B(z) &=& \displaystyle \frac{(1+(q^2-1)z)^n}{|C|} A
\left(\frac{1-z}{1+(q^2-1)z)}\right)\\
&=& \displaystyle\frac{1}{|C|} \sum_{x=0}^n A_x (1-z)^x(1+(q^2-1)z)^{n-x}.
   \end{array}
$$
We obtain the result by comparing the coefficients of $z^j$ on both sides.
\end{proof}

The theory of Shor-Laflamme weight enumerators~\cite{shor97} was
considerably extended by Rains in~\cite{rains98,rains99b,rains99d,rains00}.

\section{Bounds}
We need some bounds on the achievable minimum distance of a quantum
stabilizer code. The first theorem yields a bound that is well-suited for
computer search.
\begin{theorem}
If an $((n,K,d))_q$ stabilizer code with $K>1$ exists, then there
exists a solution to the optimization problem: minimize
$\sum_{j=1}^{d-1} A_j$ subject to the constraints
\begin{enumerate}
\item $A_0=1$ and $A_j\ge 0$ for all $1\le j\le n$;
\item  $\ds\sum_{j=0}^n A_j = q^n/K$;
\item $B_j = \ds\frac{K}{q^n} \sum_{r=0}^n K_j(r)A_r$ holds for all $j$ in the range $0\le j\le n$;
\item $A_j=B_j$ for all $j$ in $0\le j<d$ and $A_j\le B_j$ for all $d\le j\le n$;
\item $(p-1)$ divides $A_j$ for all $j$ in the range $1\le j\le n$.
\end{enumerate}
\end{theorem}
\begin{proof}
If an $((n,K,d))_q$ stabilizer code exists, then the symplectic weight
distribution of the associated additive code~$C$ satisfies
conditions~1) and~2).  For each nonzero codeword $c$ in $C$, $\alpha
c$ is again in $C$ for all $\alpha$ in $\F_p^*$, so 5)
holds. Corollary~\ref{th:krawtchouk} shows that 3) holds. Since the
quantum code has minimum distance $d$, it follows that 4) holds.  
\end{proof}

\begin{remark}
If we are interested in bounds for $\F_{q^2}$ \-linear codes, then we
can replace condition 5) in the previous theorem by $q^2-1$ divides
$A_j$.  This will even help in characteristic 2.
\end{remark}

The next bound is more convenient when one wants to find bounds by
hand. In particular, any function $f$ satisfying the constraints of
the next theorem will yield a useful bound on the dimension of a
stabilizer code. This approach was introduced by Delsarte for
classical codes~\cite{delsarte72}. Binary versions of
Theorem~\ref{th:lp2} and Corollary~\ref{th:singleton} were proved by
Ashikhmin and Litsyn~\cite{ashikhmin99}, see also~\cite{ashikhmin00b}.

\begin{theorem}\label{th:lp2}
Let $Q$ be an $((n,K,d))_q$ stabilizer code of dimension $K>1$.
Suppose that $S$ is a nonempty subset of $\{0,\dots,d-1\}$ and
$N=\{0,\dots,n\}$.  Let
$$ f(x)= \sum_{i=0}^n f_i K_i(x)$$
be a polynomial satisfying the conditions
\begin{enumerate}
\item[i)] $f_x> 0$ for all $x$ in $S$, and $f_x\ge 0$ otherwise;
\item[ii)] $f(x)\le 0$ for all $x$ in $N\setminus S$.
\end{enumerate}
Then
$$ K \le \frac{1}{q^n}\max_{x\in S} \frac{f(x)}{f_x}.$$
\end{theorem}
\begin{proof}
Suppose that $C\le \F_q^{2n}$ is the additive code associated with the
stabilizer code $Q$. If we apply Corollary~\ref{th:krawtchouk} to the
trace-symplectic dual code $C^\sdual$ of the code $C$, then we obtain
$$ A_i = \frac{1}{|C^{\sdual}|}\sum_{x=0}^n K_i(x)B_x.$$
Using this relation, we find that
$$ \begin{array}{lcl}
\ds|C^\sdual| \sum_{i\in S} f_i A_i &\le&
\ds |C^\sdual| \sum_{i=0}^{n} f_i A_i \\
&=& \ds |C^\sdual| \sum_{i=0}^{n} f_i
\left(\frac{1}{|C^{\sdual}|}\sum_{x=0}^n K_i(x)B_x\right)\\
&=& \ds \sum_{x=0}^n B_x \sum_{i=0}^n f_iK_i(x).
   \end{array}
$$
By assumption, $f(x)=\sum_{i=0}^n f_iK_i(x)$; thus,
we can simplify the latter inequality and obtain
$$
\ds|C^\sdual| \sum_{i\in S} f_i A_i \le
\ds \sum_{x=0}^n B_x f(x) \le
\sum_{x\in S} B_x f(x) = \sum_{x\in S} A_x f(x),$$
where the last equality follows from the fact that the stabilizer code
has minimum distance $d$, meaning that $A_x=B_x$ holds for all $x$ in the
range $0\le x<d$. We can conclude that
$$ |C^\sdual| \le \
\left(\ds\sum_{x\in S} A_x f(x)\right)\bigg/
\left(\ds\sum_{x\in S} f_x A_x\right)
\le \max_{x\in S} \frac{f(x)}{f_x},
$$
which proves the theorem, since $|C^\sdual|=q^n K$.
\end{proof}

As an example, we demonstrate that the previous theorem implies the
quantum Singleton bound. Linear programming yields in general better
bounds, but for short lengths one can
actually find codes meeting the quantum Singleton bound.
\begin{corollary}[Quantum Singleton Bound]\label{th:singleton}
An  $((n,K,d))_q$ stabilizer code with $K>1$ satisfies
$$ K\le q^{n-2d+2}.$$
\end{corollary}
\begin{proof}
Let $S=\{0,\dots,d-1\}$.
If we choose the polynomial
$$ f(x)=q^{n-d+1}\prod_{j=d}^n\left(1-\frac{x}{j}\right),$$ then
$f(x)=0$ for all $x$ in $\{0,\dots,n\}\setminus S$.
We can express $f(x)$
in the form
$$ f(x)=q^{n-d+1} {n-x \choose n-d+1}\bigg/ { n\choose n-d+1}.$$
We can express this polynomial as $f(x)=\sum_{i=0}^n f_iK_i(x)$, where
$$ f_i = q^{-2n} \sum_{x=0}^n f(x)K_x(i)=  q^{1-d-n}\sum_{x=0}^nK_x(i)
{n-x \choose n-d+1}\bigg/ { n\choose n-d+1}.$$
Notice that $\sum_{x=0}^n K_x(i){n-x\choose n-d+1} = {n-i\choose d-1}q^{2(d-1)}$, see~\cite{levenshtein95}; hence,
$$ f_i = q^{d-1-n} { n-i\choose d-1}\bigg/ { n\choose n-d+1}>0.$$
We obtain for the fraction
$r(x):=f(x)/f_x$ the value
$$ r(x)=\frac{f(x)}{f_x}
= q^{2n-2d+2} {n-x\choose n-d+1}\bigg/ {n-x\choose d-1}.
$$
An easy calculation shows that
$$ \frac{r(x)}{r(x+1)}= \frac{n-x-d+1}{d-x-1}.$$ Seeking a
contradiction, we assume that there exists an $((n,K,d))_q$ stabilizer
code with $2d\ge n+2$. In this case $r(x)/r(x+1)\le 1$, so that
$r(d-1)$ is the maximum of the values $r(x)$ with
$x\in\{0,\dots,d-1\}$. By Theorem~\ref{th:lp2}, we have $K\le
r(d-1)/q^n=q^{n-2d+2}/{n-d+1\choose d-1}$.  This yields a
contradiction, since ${n-d+1\choose d-1}K$ cannot be less than
$q^{n-2d+2}\le 1$ for dimension $K>1$.

If $2d < n+2$, then $r(x)/r(x+1)>1$, so $r(0)=f(0)/f_0$ is the largest
among the values $r(x)$ with $x\in \{0,\dots, d-1\}$. We have
$r(0)= q^{2n-2d+2}$; whence, it follows from
Theorem~\ref{th:lp2} that the
dimension $K$ of the code is bounded by
$$ K\le q^{-n} \max_{0\le x<d} \frac{f(x)}{f_x}=q^{n-2d+2},$$
which proves the claim.
\end{proof}
The binary version of the quantum Singleton bound was first proved by
Knill and Laflamme in~\cite{KnLa97}, see
also~\cite{ashikhmin99,ashikhmin00b}, and later generalized by Rains
using weight enumerators in~\cite{rains99}.

The quantum Hamming bound states that any pure $((n,K,d))_q$
stabilizer code satisfies
$$ \sum_{i=0}^{\lfloor (d-1)/2\rfloor} \binom{n}{i}(q^2-1)^i \le
q^n/K,$$ see~\cite{gottesman96,feng04}. Several researchers have tried
to find impure stabilizer codes that beat the quantum Hamming bound.
However, Gottesman has shown that impure single and double
error-correcting binary quantum codes cannot beat the quantum Hamming
bound~\cite{gottesman97}.  In the same vein, Theorem~\ref{th:lp2}
allows us to derive the Hamming bound for arbitrary stabilizer codes,
at least when the minimum distance is small. We illustrate the method
for single error-correcting codes, and note that the same approach
works for double error-correcting codes as well.

\begin{corollary}[Quantum Hamming Bound]\label{th:hamming}
An $((n,K,3))_q$ stabilizer code with $K>1$ satisfies
$$ K\le q^{n}\big/(n(q^2-1)+1).$$
\end{corollary}
\begin{proof}
Recall that the intersection number $p_{ij}^k$ of the Hamming
association scheme $H(n,q^2)$ is the integer
$ p_{ij}^k = | \{ z\in \F_{q^2}^n \,|\, d(x,z)=i, d(y,z)=j\}|,$
where $x$ and $y$ are two vectors in $\F_q^n$ of Hamming
distance~$d(x,y)=k$. The intersection numbers are related to
Krawtchouk polynomials by the expression
$$ p_{ij}^k = q^{-2n} \sum_{u=0}^n K_i^n(u)K_j^n(u)K_u^n(k),$$
see~\cite{barg00}.

After this preparation, we can proceed to derive the Hamming bound as
a consequence of Theorem~\ref{th:lp2}. Let
$$ f(x) = \sum_{j,k=0}^1 \sum_{i=0}^n K_j^n(i)K_k^n(i)K_i^n(x) =
q^{2n}(p_{00}^x+ p_{10}^x+ p_{01}^x + p_{11}^x).$$ The triangle
inequality implies that $p_{ij}^k=0$ if one of the three arguments
exceeds the sum of the other two; hence, $f(x)=0$ for $x>2$.
The coefficients of the Krawtchouk expansion
$f(x)=\sum_{i=0}^n f_i K_i(x)$ obviously satisfy $f_i=(K_0(i)+K_1(i))^2\ge 0$.
A straightforward
calculation gives
$$
\begin{array}{lcl@{\quad\qquad}lcl}
f(0)&=& q^{2n}(n(q^2-1)+1), & f_0 &=& (n(q^2-1)+1)^2,\\
f(1)&=& q^{2n+2},           & f_1 &=& ((n-1)(q^2-1))^2,\\
f(2)&=& 2q^{2n},            & f_2 &=& ((n-2)(q^2-1)-1)^2.
\end{array}
$$
It follows that
$$ \max\{ f(0)/f_0, f(1)/f_1, f(2)/f_2\} \le q^{2n}/(n(q^2-1)+1)$$
holds for all $n\ge 5$. Using Theorem~\ref{th:lp2}, we obtain the
claim for all $n\ge 5$. For the lengths $n<5$, we obtain the claim
from the quantum Singleton bound.
\end{proof}

One real disadvantage of Theorem~\ref{th:lp2} is that the number of
terms increase with the minimum distance and this can lead to
cumbersome calculations. However, one can derive more consequences
from Theorem~\ref{th:lp2}; see, for
instance,~\cite{ashikhmin99,ashikhmin00b,levenshtein95,mceliece77}.

\paragraph{Lower Bounds.}
We conclude this section by giving the quantum version of the
classical lower bounds by Gilbert and Varshamov. Basically, a simple
counting argument is used to establish the existence of stabilizer
codes.  

Our first lemma generalizes an idea used by Gottesman in his
proof of the binary case.
\begin{lemma}\label{th:gilbert}
An $((n,K,\ge d))_q$ stabilizer code with $K>1$ exists provided that
\begin{equation}\label{eq:gilbert}
(q^nK-q^n/K) \sum_{j=1}^{d-1} \binom{n}{j}(q^{2}-1)^j
<(q^{2n}-1)(p-1)
\end{equation}
holds.
\end{lemma}
\begin{proof}
Let $L$ denote the multiset
$$L=\{ C^\sdual\setminus C\,|\, C\le C^\sdual\le \F_q^{2n} \text{ with
} |C|=q^n/K\}.$$ The elements of this multiset correspond to
stabilizer codes of dimension $K$.  Note that $L$ is nonempty, since
there exists a code $C$ of size $q^n/K$ that is generated by elements
of the form $(a|0)$; the form of the generators ensures that $C\le
C^\sdual$.

All nonzero vectors in $\F_q^{2n}$ appear in the same number of sets
in $L$. Indeed, the symplectic group $\textup{Sp}(2n,\F_q)$ acts
transitively on the set $\F_{q}^{2n}\setminus \{ 0\}$,
see~\cite[Proposition~3.2]{grove01}, which means that for any nonzero
vectors $u$ and $v$ in $\F_q^{2n}$ there exists $\tau\in
\textup{Sp}(2n,\F_q)$ such that $v=\tau u$. Therefore, $u$ is
contained in $C^\sdual\setminus C$ if and only if $v$ is contained in
the element $(\tau C)^\sdual\setminus \tau C$ of $L$.

The transitivity argument shows that any nonzero vector in
$\F_{q}^{2n}$ occurs in $|L|(q^nK-q^n/K)/(q^{2n}-1)$ elements of $L$.
Furthermore, a nonzero vector and its $\F_p^\times$-multiples are contained
in the exact same sets of $L$.
Thus, if we delete all sets from $L$
that contain a nonzero vector with symplectic weight less than~$d$,
then we remove at most
$$
\frac{\sum_{j=1}^{d-1} \binom{n}{j}(q^{2}-1)^j}{p-1}
|L|\frac{(q^nK-q^n/K)}{q^{2n}-1}$$ sets from $L$. By assumption, this
number is less than $|L|$; hence, there exists an $((n,K,\ge d))_q$
stabilizer code.
\end{proof}

The Gilbert-Varshamov bound provides surprisingly good codes, even for
smaller lengths, when the characteristic of the field is
not too small. If $n\equiv k\bmod 2$, then we can significantly
strengthen the bound. 

\begin{lemma}\label{th:lingilbert}
If $k\ge 1$, $n\equiv k\bmod 2$ and
\begin{equation}\label{eq:lingilbert}
(q^{n+k}-q^{n-k}) \sum_{j=1}^{d-1} \binom{n}{j}(q^{2}-1)^{j-1}
<(q^{2n}-1)
\end{equation}
holds, then there exists an $\F_{q^2}$-linear $[[n,k,d]]_{q}$ stabilizer code.
\end{lemma}
\begin{proof}
The proof is almost the same as in the previous lemma, except that we
list only codes $C$ that are linear, meaning that $\phi(C)$ is a
vector space over $\F_{q^2}$. We repeat the previous argument with the
multiset
$$ L = \{ C^\sdual\setminus C\,|\, C\le C^\sdual \le F_q^n,
|C|=q^{n-k}, \phi(C) \text{ is $\F_{q^2}$-linear } \}. $$ Note that
each set $C^\sdual \setminus C$ in $L$ contains now all
$\F_{q^2}^\times$-multiples of a nonzero vector, not just the
$\F_p^\times$-multiples, which proves the statement.
\end{proof}

Feng and Ma have recently shown that one can extend the previous
result to even prove the existence of pure stabilizer codes, but much
more delicate counting arguments are needed in that case,
see~\cite{feng04}. We are not aware of short proofs for this stronger
result.

The previous lemma allows us to derive good quantum codes, especially
for larger alphabets. We illustrate this fact by proving the existence
of MDS stabilizer codes, see Section~\ref{sec:MDS} for more details on
such codes.

\begin{corollary}
If $2\le d\le \lceil n/2\rceil $ and $q^2-1\ge \binom{n}{d}$, then
there exists a linear $[[n,n-2d+2,d]]_q$ stabilizer code.
\end{corollary}
\begin{proof}
The assumption $d\le \lceil n/2\rceil $ implies that $\binom{n}{1}\le
\binom{n}{2}\le \cdots \le \binom{n}{d}$, so the maximum value of
these binomial coefficients is at most $q^2-1$. Let $k=n-2d+2$. It
follows from the assumption that $k\ge 1$ and $n\equiv k\bmod
2$. It remains to show that (\ref{eq:lingilbert}) holds.
For the choice $k=n-2d+2$,
the left hand side of (\ref{eq:lingilbert}) equals
\begin{equation*}
\begin{split}
\ds(q^{2n-2d+2}-q^{2d-2})\sum_{j=1}^{d-1} \binom{n}{j}&(q^2-1)^{j-1}\le
\ds (q^{2n-2d+2}-q^{2d-2})\sum_{j=1}^{d-1} (q^2-1)^{j}\\
&=\ds(q^{2n-2d+2}-q^{2d-2}) \frac{(q^2-1)^d-(q^2-1)}{q^2-2}.
\end{split}
\end{equation*}
We claim that the latter term is less than $q^{2n}-1$.
To prove this, it suffices to show that
\begin{equation}\label{eq:ineq}
q^{2n-2d+2}\frac{(q^2-1)^d-(q^2-1)}{q^2-2}\le q^{2n}
\end{equation}
holds. The latter inequality is equivalent to $ (q^2-1)^d \le
q^{2d}-2q^{2d-2}+q^2-1$, and it is not hard to see that this
inequality holds.  Indeed, note that
$$ q^{2d}=((q^2-1)+1)^d = (q^2-1)^d+ \sum_{j=0}^{d-1} \binom{d}{j}(q^2-1)^j.$$\
Recall that $\binom{d}{j}=\binom{d-1}{j-1}+\binom{d-1}{j}$; hence,
\begin{eqnarray*} 
q^{2d}-2q^{2d-2}- (q^2-1)^d &=& \sum_{j=0}^{d-1}\big(
\binom{d}{j}-2\binom{d-1}{j} \big)(q^2-1)^j,\\
&=& \sum_{j=0}^{d-1}\big(
\underbrace{\binom{d-1}{j-1}-\binom{d-1}{j}}_{\alpha(j):=}\big)(q^2-1)^j.
\end{eqnarray*}
We have $\alpha(j)=-\alpha(d-j)$ for $0\le j\le d-1$, and
$\alpha(j)\ge 0$ for $j\ge d/2$. This shows that all negative terms
get canceled by larger positive terms and we can conclude that
$q^{2d}-2q^{2d-2}-(q^2-1)^d\ge 0$ for $d\ge 2$; this implies
inequality~(\ref{eq:ineq}) and consequently shows
that~(\ref{eq:lingilbert}) holds.
\end{proof}

\begin{example} 
Recall that there does not exist a $[[7,1,4]]_2$ code,
see~\cite{calderbank98}. In contrast, the existence of a $[[7,1,4]]_q$
code for all prime powers $q\ge 7$ is guaranteed by the preceding
corollary. It also shows that there exist $[[6,2,3]]_q$ for all prime
powers $q\ge 5$ and $[[7,3,3]]_q$ for all prime powers $q\ge 7$, which
slightly generalizes~\cite{feng02}.
\end{example}


\section{Cyclic Codes}
We shall now restrict our attention to linear quantum codes and derive 
several families of quantum codes from classical linear codes. In essence 
we make use of the hermitian and CSS constructions (Lemmas 
\ref{co:classical}~-~\ref{th:css2}). Hence, we need to look for classical codes
that are self-orthogonal with respect to the hermitian or
the euclidean product or families of nested codes like the BCH codes.
 
In case of cyclic codes identifying the self-orthogonal codes can be
translated into equivalent conditions on the generator polynomial 
of the code or its defining set. 
Let $\sigma$ denote the automorphism of the field $\F_{q^2}$ given by
$\sigma(x)=x^q$. We can define an action of $\sigma$ on the polynomial
ring $\F_{q^2}[x]$ by
$$h(x)=\sum_{k=0}^n h_k x^k \longmapsto h^\sigma(x)
=\sum_{k=0}^n \sigma(h_k) x^k.$$

\begin{lemma}\label{th:cyclic}
Suppose that $B$ is a classical cyclic $[n,k,d]_{q^2}$ code with
generator polynomial $g(x)$ and check polynomial
$h(x)=(x^n-1)/g(x)$. If $g(x)$ divides $\sigma(h_0)^{-1}x^k
h^\sigma(1/x)$, then $B^\hdual\subseteq B$, and there exists an
$[[n,2k-n,\geq d]]_{q}$ stabilizer code that is pure to~$d$.
\end{lemma}
\begin{proof}
If $h(x)$ is the check polynomial of $B$, then $h^\sigma(x)$ is the
check polynomial of $\sigma(B)$. The generator polynomial of the dual
code $\sigma(B)^\perp=B^\hdual$ is given by $\sigma(h_0)^{-1}x^k
h^\sigma(1/x)$, the normalized reciprocal polynomial of $h^\sigma(x)$.
Therefore, the condition that the polynomial $g(x)$ divides
$\sigma(h_0)^{-1}x^k h^\sigma(1/x)$ is equivalent to the condition
$B^\hdual \subseteq B$. The stabilizer code follows from
Corollary~\ref{co:classical}.
\end{proof}

The polynomial $x^n-1$ of $\F_{q^2}[x]$ has simple roots if and only
if $n$ and $q$ are coprime. If the latter condition is satisfied,
then there exists a positive integer $m$ such that the field
$\F_{q^{2m}}$ contains a primitive $n$th root of unity $\beta$.  In
that case, one can describe a cyclic code with generator polynomial
$g(x)$ in terms of its defining set $Z=\{ k\,|\, g(\beta^k)=0 \text{ for
} 0\le k< n\}$. The following Lemma summarizes various equivalent conditions on
self-orthogonal codes in terms of the generator polynomial and the
defining set $Z$.

\begin{lemma}\label{th:cyclic2}
Let $\gcd(n,q^2)=1$ and $C$ be a classical cyclic $[n,k,d]_{q^2}$ code
whose generator polynomial is $g(x)$ and defining set $Z$. If any of
the following equivalent conditions are satisfied\\ (i) $x^n - 1
\equiv 0 \mod g(x)g^{*}(x)$ where $g^{*}(x)=x^{n-k}g^\sigma(1/x)$\\
(ii) $Z\subseteq \{ -qz\,|\, z\in N\setminus Z\}$\\ (iii) $Z\cap
Z^{-q}=\emptyset $, where $Z^{-q}=\{-qz\mid z\in Z\}$\\ then $C^\hdual
\subseteq C $ and there exists an $[[n,2k-n,\geq d]]_{q}$ stabilizer
code that is pure to~$d$.
\end{lemma}

\begin{proof}
Let $h(x)=(x^n-1)/g(x)$ be the check polynomial of $C$. 
Then $h^\sigma(x)=\sigma((x^n-1)/g(x))=(x^n-1)/g^\sigma(x)$. 
From Lemma~\ref{th:cyclic} we know that $C$ contains its hermitian dual if 
g(x) divides $\sigma(h_0)^{-1}x^k h^\sigma(1/x)$ viz. 
$g(x) | \sigma(h_0)^{-1}(1-x^n)/(x^{n-k}g^\sigma(1/x))$, which 
implies $x^n-1 \equiv 0 \mod g(x) g^{*}(x)$ which proves (i).

The generator polynomial $g(x)$ of $C$ is given by
$g(x)=\prod_{z\in Z} (x-\beta^z)$, hence its check polynomial is
of the form
$$h(x)=(x^n-1)/g(x)=\prod_{z\in N\setminus Z} (x-\beta^z).$$
Applying the automorphism $\sigma$ yields
$h^\sigma(x)= \prod_{z\in N\setminus Z} (x-\beta^{qz}).$
Therefore, the generator polynomial of $C^\hdual$ is given by
$$ \begin{array}{lcl}
h^\sigma(0)^{-1} x^k h^\sigma(1/x)
&=& h^\sigma(0)^{-1} \prod_{z\in N\setminus Z} (1-\beta^{qz}x)\\
&=& \prod_{z \in N\setminus Z} (x -\beta^{-qz});
   \end{array}
$$
in the last equality, we have used the fact that
$h^\sigma(0)^{-1}=\prod_{z\in N\setminus Z} (-\beta^{-qz})$. By Lemma~\ref{th:cyclic}, $B^\hdual \subseteq B$ if
and only if the generator polynomial $g(x)$ divides $h^\sigma(0)^{-1}
x^k h^\sigma(1/x)$. The latter condition is equivalent to the fact
that $Z$ is a subset of $\{ -qz\mid z\in N\setminus Z\}$ and (ii) follows.
From (ii)  we know that $C^\hdual \subseteq C$ if and only if $Z\subseteq \{ -qz\mid z\in
N\setminus Z\}$. In other words $Z^{-q} \subseteq N\setminus Z$.
Hence $Z\cap Z^{-q}=\emptyset$. 
An  $[[n,2k-n,\geq d]]_q$ stabilizer code follows from Corollary~\ref{co:classical}.
\end{proof}

Cyclic codes that contain their euclidean duals can also be nicely
characterized in terms of their generator polynomials and defining sets. 
The following Lemma is a very straight forward extension of the binary 
case and summarizes some of the known results in the nonbinary case
as well, but we include it because of its usefulness in
constructing cyclic quantum codes. 

\begin{lemma}\label{th:csscyclic}
Let $C$ be an $[n,k,d]_q$ cyclic code such that $\gcd(n,q)=1$. Let its defining set $Z$ and
generator polynomial  $g(x)$ be such that any of the following
equivalent conditions are satisfied\\
(i) $x^n-1 \equiv 0 \mod g(x)g^{\dagger}(x)$, where 
$g^{\dagger}(x)=x^{n-k}g(1/x)$; \\
(ii) $Z \subseteq \{-z\mid z\in N\setminus Z\}$; \\
 (iii) $Z\cap Z^{-1}=\emptyset$ where $Z^{-1}=\{ -z\mod n\mid z\in Z\}$.\\
Then $C^\perp \subseteq C$ and there exists an $[[n,2k-n,\geq d]]_q$
stabilizer code that is pure to~$d$.
\end{lemma}
\begin{proof}
The check polynomial of $C$ is given by $h(x)=(x^n-1)/g(x)$, from
which we obtain the (un-normalized) generator polynomial of $C^\perp$
as
$h^\dagger(x)=x^kh(x^{-1})=(1-x^n)/(x^{n-k}g(x^{-1}))=-(x^n-1)/g^\dagger(x)$.
If $C^\perp \subseteq C$, then $g(x)\mid h^\dagger(x)$; this means
that $g(x)$ divides $(x^n-1)/g^\dagger(x)$. In other words $ x^n-1
\equiv 0\mod g(x)g^\dagger(x)$.

The defining set of $C^\perp$ is given by $\{-z\mod n \mid z\in
N\setminus Z \}$, where $N=\{0,1,\ldots,n-1 \}$. Thus, $C^\perp
\subseteq C$ implies $Z\subseteq \{ -z\mod n\mid N\setminus Z\}$.
Since this means that the inverses of elements in $Z$ are present in 
$N\setminus Z$, this condition can also be written as $Z\cap
Z^{-1}=\emptyset$. The
existence of  quantum code $[[n,2k-n,\geq d]]_q$ follows from Corollary
\ref{th:css2}.
\end{proof}

Although we have considered purely cyclic codes, a larger class of
cyclic quantum codes can be derived by considering constacyclic or
conjucyclic codes as in \cite{calderbank98}, \cite{xiaoyan04}.

\section{Cyclic Hamming Codes}
Binary quantum Hamming codes have been studied by various authors; see
for instance \cite{gottesman96,calderbank98,feng02b}.
We will now derive stabilizer codes from nonbinary classical cyclic Hamming
codes. Let $m>1$ be an integer such that
$\gcd(q-1,m)=1$. A classical cyclic Hamming code $H_q(m)$ has
parameters $[n,n-m,3]_{q}$ with length $n=(q^{m}-1)/(q-1)$.
Let $\beta$ denote a primitive
$n$th root of unity in $\F_{q^{m}}$. The generator polynomial of
$H_q(m)$ is given by
\begin{eqnarray}
 g(x)=\prod_{i=0}^{m-1} \big(x-\beta^{q^{i}}\big), \label{eq:hamminggen}
\end{eqnarray}
an element of
$\F_{q}[x]$. Thus, the code $H_q(m)$ is defined by the cyclotomic coset
$C_1=\{ q^{i}\bmod n \,|\, i \in \Z\}$.

\begin{lemma}\label{th:hammingdual}
The Hamming code $H_{q^2}(m)$ contains its hermitian dual, that is, 
$H_{q^2}(m)^\hdual\le H_{q^2}(m)$.
\end{lemma}
\begin{proof}
The statement $H_{q^2}(m)^\hdual\le H_{q^2}(m)$ is equivalent to the fact that the
cyclotomic coset $C_1$ satisfies $C_1\subseteq N_1=\{ -qz\bmod n\,|\,
z\in N\setminus C_1\}$, where $N=\{0,\dots,n-1\}$ and $n=(q^{2m}-1)/(q^2-1)$.  We note that $C_1$
can be expressed in the form
\begin{equation}\label{eq:cosets}
C_1=\left\{ (1-n)q^{2k}\bmod n \,\Big|\, k\in \Z\right\}=
\left\{ -qzq^{2k} \bmod n\,\Big|\, k\in \Z\right\},
\end{equation}
where $z=q(q^{2m-2}-1)/(q^2-1)$.  Therefore, the condition
$C_1\subseteq N_1$ holds if and only if $C_z\subseteq N\setminus C_1$ holds, where $C_z=\{zq^{2j}\bmod n\,|\, j\in
\Z\}$.

Seeking a contradiction, we assume that the two cyclotomic cosets
$C_1$ and $C_z$ have an element in common, hence are the same.  This
means that there must exist a positive integer $k$ such that
$q^{2k}=q(q^{2m-2}-1)/(q^2-1)$. This implies that
$q^{2k-1}$ divides $q^{2m-2}-1$, which is absurd.
Thus, the sets $C_1$ and $C_z$ are disjoint, hence $C_z\subseteq
N\setminus C_1$, which proves the claim.
\end{proof}

\begin{theorem}
For each integer $m\ge 2$ such that $\gcd(m,q^2-1)=1$, there exists a
pure $[[n,n-2m,3]]_q$ stabilizer code of length
$n=(q^{2m}-1)/(q^2-1)$.
\end{theorem}
\begin{proof}
If $\gcd(m,q^2-1)=1$, then there exists a classical $[n,n-m,3]_{q^2}$
Hamming code $H_{q^2}(m)$. By Lemma~\ref{th:hammingdual}, we have
$H_{q^2}(m)^\hdual\le H_{q^2}(m)$, hence there exists a pure $[[n,n-2m,3]]_{q}$
stabilizer code by Corollary~\ref{co:classical}. The purity is due to the fact 
that the $H_{q^2}(m)^\hdual$ has minimum distance $q^{2m-2} \geq 3$ for $m\geq 2$ \cite[Theorem~1.8.3]{huffman03}.
\end{proof}

These quantum Hamming codes are optimal since they attain the quantum
Hamming bound, see Corollary~\ref{th:hamming}. A different approach
that allows construction of noncyclic perfect quantum codes can be
found in \cite{bierbrauer00}.  It is also possible to construct
quantum codes from Hamming codes that contain their euclidean duals,
however these codes do not meet the quantum Hamming bound.

\begin{lemma}
If $\gcd(m,q-1)=1$ and $m\geq 2$, then there exists a pure
$[[n,n-2m,3]]_q$ quantum code, where $n=(q^m-1)/(q-1)$.
\end{lemma}
\begin{proof}
The generating polynomial of $[n,n-m,3]_q$ Hamming code, with
n=$(q^m-1)/(q-1)$ is given by equation~(\ref{eq:hamminggen}) where
$\beta$ is an element of order $n$.  The code exists only if
gcd$(m,q-1)=1$.  By Lemma \ref{th:csscyclic} a cyclic code contains
its dual if $x^n-1 \equiv 0 \mod g(x)g^\dagger(x)$, where
$g^\dagger(x)=x^{n-k}g(x^{-1})$. If $g(x)$ is not self-reciprocal then
$g(x)g^\dagger(x)$ divides $x^n-1$ \cite{vatan99}. Since the
generating polynomial of the Hamming code is not self-reciprocal, the
code contains its euclidean dual. By Lemma \ref{th:csscyclic} we can
construct a quantum code with the parameters $[[n,n-2m,3]]_q$.  Once
again the purity follows due to the fact the duals of Hamming codes
are simplex codes with weight $q^{m-1} \geq 3$
\cite[Theorem~1.8.3]{huffman03} for $m\geq 2$.
\end{proof}

\section{Quadratic Residue Codes}

Another well known family of classical codes are the quadratic residue codes.
Rains constructed quadratic residue codes for prime alphabet in \cite{rains99}.
In this section we will construct two series of quantum codes based on the
classical quadratic codes for any arbitrary field using elementary methods. 

\begin{theorem}[Quadratic Residue Codes]
Let $n$ be a prime of the form $n\equiv 3 \mod 4$, and let
$q$ be a power of a prime that is not divisible by~$n$.
If\/ $q$ is a quadratic residue modulo~$n$, then there exists
a pure $[[n,1,d]]_q$ stabilizer code with minimum distance $d$ satisfying
$d^2-d+1\ge n$.
\end{theorem}
\begin{proof}
Let $\alpha$ denote a primitive $n$th root of unity from some
extension field of $\F_q$. Let $R=\{ r^2\bmod n\mid
r\in \Z \text{ such that } 1\le r\le (n-1)/2\}$ denote
the set of quadratic residues modulo
$n$. We define the quadratic residue code
$C_R$ as the cyclic code of length $n$ over $\F_q$
that is generated by the polynomial
$$ q(x) = \prod_{r\in R} (x-\alpha^r).$$ The code $C_R$ has parameters
$[n,(n+1)/2,d]_q$ and if $n\equiv 3 \bmod 4$, the dual code $C^\perp_R$ of $C_R$ is
given by the cyclic code generated by $(x-1)q(x)$, the even-like
subcode of $C_R$. The minimum distance $d$ is bounded by
$d^2-d+1\ge n$, see, for instance, \cite[pp.~114-119]{betten98}. Further 
$\wt(C_R\setminus C_R^\perp)=\wt(C_R)=d$ by \cite[Theorem~6.6.22]{huffman03}. 
We can deduce from Corollary~\ref{th:css2} that
there exists a pure $[[n,(n+1)-n,d]]_q$ stabilizer code.
\end{proof}

For example, the prime $p=3$ is a quadratic residue modulo $n=23$. The
previous proposition guarantees the existence of a $[[23,1,d]]_3$
stabilizer code with minimum distance $d\ge 6$.

If $n$ is an odd prime of the form $n\equiv 1 \bmod 4$, then we can
also construct quadratic residue codes, but now we need to
employ Lemma~\ref{th:css}, because $C_R$ does not contain
its dual.

\begin{theorem}
Let $n$ be a prime of the form $n\equiv 1 \mod 4$. Let $q$ be a power
of a prime that is not divisible by $n$. If\/ $q$ is a quadratic
residue modulo~$n$, then there exists a pure $[[n,1,d]]_q$ stabilizer code
with minimum distance $d$ bounded from below by $d\ge \sqrt{n}$.
\end{theorem}
\begin{proof}
Let $\alpha$ denote a primitive $n$th root of unity from some
extension field of $\F_q$. We denote by $R$ denote the set of
quadratic residues modulo $n$ and by $N$ the set of quadratic
non-residues modulo $n$.

Let $C_R$ and $C_N$ denote the cyclic
codes of length $n$ that are respectively generated by the polynomials
$q_R(x)$ and $q_N(x)$, where
$$ q_R(x) = \prod_{r\in R} (x-\alpha^r)\quad\mbox{and}\quad q_N(x)=
\prod_{r\in N} (x-\alpha^r).$$ Both codes have parameters
$[n,(n+1)/2,d]_q$ with $d^2\ge n$, see \cite[pp.~114-119]{betten98}.

The dual code of $C_R$ is given by the even-like subcode of $C_N$; in
other words, $C_R^\perp$ is a cyclic code of length $n$ over $\F_q$
that is generated by the polynomial $(x-1)q_N(x)$; in particular,
$C_R^\perp \le C_N$. 
Moreover $\wt(C_R\setminus C_N^\perp)=\wt(C_N\setminus C_R^\perp)=d$ by \cite[Theorem~6.6.22]{huffman03}.
Therefore, we obtain a pure $[[n,(n+1)/2+(n+1)/2-n,d]]_q$ code by Lemma~\ref{th:css}.
\end{proof}

\section{Quantum  Melas  Codes}
One of the earliest family of codes that were constructed with a view
to correcting burst errors are the Melas codes. While not as well
known as the Hamming codes or the quadratic residue codes, they are
nonetheless an interesting family of codes. These codes have been well
investigated, especially in the mathematical community, because of
their connections to algebraic geometry \cite{schoof95,geer94,lachaud90,schoof92}.  
See \cite{hiramatsu03} for an interesting read on the connections to 
number theory.

The Melas code\footnote{The classical Melas codes are defined over a
prime field $\F_p$ and have the parameters $[p^m-1,p^m-m-1,\geq
3]_p$ (cf. \cite{stichtenoth94}); here we consider a generalization to arbitrary finite
fields.}  $\mathcal{M}_{q}(m)$ is a cyclic $[n,n-2m,\geq 3]_{q}$
code with $n=q^{m}-1$. The generator polynomial of
$\mathcal{M}_{q}(m)$ is given by
\begin{eqnarray}
g(x)=\prod_{i=0}^{m-1}(x-\alpha^{q^{i}})(x-\alpha^{-q^{i}}),
\end{eqnarray}
where $\alpha$ is a primitive element in $\F_{q^{m}}$. Alternatively, 
the defining set of the code is given by $Z=C_1\cup C_{-1} =\{\pm
q^{i} \mod n \mid 0\le i<m \}$.

\begin{lemma}\label{th:melas_orth}
The Melas code $\mathcal{M}_{q^2}(m)$  contains its hermitian dual.
\end{lemma}
\begin{proof}
By Lemma \ref{th:cyclic2}, it suffices to show that $Z \cap
Z^{-q}=\emptyset$. Seeking a contradiction, we assume that $Z \cap
Z^{-q} \neq \emptyset$. Since $\gcd(q^2,q^{2m}-1)=1$, this implies
that there must exist some integer $i$ in the range $0\le i<m$ such
that $q^{2i}\equiv \pm q\mod n$, but that is impossible; 
so $Z\cap Z^{-q}=\emptyset$. 
\end{proof}

\begin{lemma}\label{th:melas_dist}
If $q$ is even, then the minimum distance of the Melas code
$\mathcal{M}_{q^2}(m)$ is at least~3.
\end{lemma}
\begin{proof}
The parity check matrix of $\mathcal{M}_{q^2}(m)$ is given by \
\begin{eqnarray}
H&=&\left( \begin{array}{ccccc}
1&\alpha&\alpha^2&\cdots&\alpha^{n-1}\\
1&\alpha^{-1}&\alpha^{-2}&\cdots&\alpha^{-(n-1)}\end{array}\right).
\label{def_CM }
\end{eqnarray}
This matrix has a rank  $2$ only if no two columns 
are scalar multiples of each other. 
Seeking a contradiction, we suppose that 
\begin{eqnarray}
\left(\begin{array}{c}\alpha^{x} \\ \alpha^{-x}\end{array} \right)
=\alpha^{t} \left(\begin{array}{c}\alpha^{y} \\
\alpha^{-y}\end{array}\right)
\end{eqnarray}
holds for distinct $x$ and $y$. This yields $\alpha^{2t}=1$, which
implies $t\in \{0,n/2\}$. If $q$ is even, then $n$ is odd, and so $t$
cannot equal $n/2$. If $t=0$, then $x=y$ contradicting the
distinctness of $x$ and $y$. Therefore, we can conclude that $H$ has
rank $r=2$; thus, the minimum distance is at least~3.
\end{proof}

\begin{theorem}[Quantum Melas codes] If $q$ is even and $n=q^{2m}-1$,
there exist quantum Melas codes with parameters $[[n,n-4m,\ge
3]]_q$ that is pure to 3.
\end{theorem}
\begin{proof}
By Lemma \ref{th:melas_orth} we have $\mathcal{M}_{q^2}(m)^\hdual
\subseteq \mathcal{M}_{q^2}(m)$ and by Lemma \ref{th:melas_dist} we
have the distance $\geq 3$. So by Corollary \ref{co:classical} there
exists an $[[n,n-4m,\geq 3]]_q$ quantum code.
\end{proof}

\section{Quantum BCH Codes}
In this section we consider a popular family of classical codes, the
BCH codes, and construct the associated nonbinary quantum stabilizer
codes.  Binary quantum BCH codes were studied in
\cite{calderbank98,cohen99,grassl99b, steane99}. The CSS construction
turns out to be especially useful, because BCH codes form a naturally
nested family of codes.  In case of primitive BCH codes over prime
fields the distance of the dual is lower bounded by the generalized
Carlitz-Uchiyama bound, and this allows us to derive bounds on the
minimum distance of the resulting quantum codes.

Let $q$ be a power of a prime and $n$ a positive integer that is
coprime to~$q$. Recall that a BCH code $C$ of length $n$ and designed
distance $\delta$ over $\F_q$ is a cyclic code whose defining set $Z$
is given by a union of $\delta-1$ subsequent cyclotomic cosets,
$$Z=\bigcup_{x=b}^{b+\delta-2} C_x, \quad\text{where} \quad C_x =
\{xq^{r} \bmod n \mid r \in \Z, r\ge 0 \}.$$ 
The generator polynomial of the code is of the form 
$$ g(x) = \prod_{z \in Z} (x-\beta^{z}),$$ where $\beta$ is a primitive
$n$-th root of unity of some extension field of $\F_q$.  The
definition ensures that $g(x)$ generates a cyclic $[n,k,d]_{q}$ code
of dimension $k=n-|Z|$ and minimum distance~$d\ge \delta$.
If $b=1$, then the code $C$ is called a narrow-sense BCH code, and if
$n=q^m-1$ for some $m\ge 1$, then the code is called primitive.

\paragraph{Generalized Carlitz-Uchiyama Bound.} 
Our first construction derives stabilizer codes from BCH codes over
prime fields. We use the Knuth-Iverson bracket $[statement]$ in the
formulation of the Carlitz-Uchiyama bound that evaluates to 1 if
$statement$ is true and 0 otherwise.

\begin{lemma}[Generalized Carlitz-Uchiyama Bound]\label{th:carlitz}
Let $p$ be a prime. 
Let $C$ denote a narrow-sense BCH code of length $n=p^m-1$ over $\F_p$,
of designed distance $\delta=2t+1$. Then the minimum
distance $d^\perp$ of its euclidean dual code $C^\perp$ is bounded by 
\begin{equation}\label{eq:carlitzdist} 
d^\perp \ge  \Big(1-\frac{1}{p}\Big)
\left( p^m-\frac{\delta-2-[\delta-1\equiv 0\bmod p]}{2}
\big\lfloor 2p^{m/2}\big\rfloor\right).
\end{equation}
\end{lemma}
\begin{proof}
See \cite[Theorem 7]{stichtenoth94}; for further background,
see~\cite[page 280]{macwilliams77}. 
\end{proof}

\begin{theorem} 
Let $p$ be a prime. Let $C$ be a $[p^m-1,k,\ge \delta]_p$ narrow-sense
BCH code of designed distance $\delta=2t+1$ and $C^*$ a
$[p^m-1,k^*,d^*]_p$ BCH code such that $C\subseteq C^*$. Then there
exists a $[[p^m-1,k^*-k,\ge \min\{d^*,d^\perp\}]]_p$ stabilizer code,
where $d^\perp$ is given by (\ref{eq:carlitzdist}).
\end{theorem}
\begin{proof}
The result follows from applying Lemma~\ref{th:carlitz} to $C$ and
Lemma~\ref{th:css} to the codes $C$ and $C^*$. 
\end{proof}
\begin{remark} \begin{inparaenum}[(i)]
\item The Carlitz-Uchiyama bound becomes trivial for larger design distances.
\item In \cite[Corollary 2]{moreno94} it was shown that for binary BCH
codes of design distance $d$, the lower bound in equation
(\ref{eq:carlitzdist}) is attained when $n=2^{2ab}-1$, where $a$ is
the smallest integer such that $d-2\mid 2^{a}+1$ and $b$ is odd.
\item For a further tightening of the Carlitz-Uchiyama bound see
\cite[Theorem 2]{moreno98}.  \end{inparaenum}
\end{remark}

\paragraph{Primitive BCH codes containing their duals.}
We can extend result of the previous section to BCH codes over finite
fields that are not necessarily prime. In fact, if we restrict
ourselves to smaller designed distances, then we can even achieve
significantly sharper results.\nocite{steane99} We will just review
the results and refer the reader to our companion
paper~\cite{preprint0501126} for the proofs.

In the BCH code construction, it is in general not obvious how large the
cyclotomic cosets will be. However, if the designed distance is small,
then one can show that the cyclotomic cosets all have maximal size. 

\begin{lemma}\label{th:bchdim} 
A narrow-sense, primitive BCH code with design distance 
$2\leq \delta \leq q^{\lceil m/2\rceil}+1$ has parameters
$[q^m-1,q^m-1-m\lceil (\delta-1)(1-1/q) \rceil,\geq \delta]_q$. 
\end{lemma}
\begin{proof}
See \cite[Theorem A]{preprint0501126}; the binary case was already
established by Steane~\cite{steane99}.
\end{proof}

In the case of small designed distances, primitive, narrow-sense 
BCH codes contain their euclidean duals. 
\begin{lemma}\label{th:bcheuclideandual}

A narrow-sense, primitive BCH code over $\F_q^n$ contains its
euclidean dual if and only if its design
distance $2 \leq \delta \leq q^{\lceil m/2\rceil}-1-(q-2)[m \textrm{
odd}]$, where $n=q^m-1$ and $m\geq 2$.
\end{lemma}
\begin{proof}
See \cite[Theorem C]{preprint0501126}. 
\end{proof}

A simple consequence is the following theorem: 

\begin{theorem}\label{co:bcheuclideandual}
If $C$ is a narrow-sense primitive BCH code over $\F_q$ with design
distance $2\leq \delta \leq q^{\lceil m/2 \rceil}-1-(q-2)[m \textrm{
odd}]$ and $m\geq 2$, then there exists an
$[[q^m-1,q^m-1-2m\lceil(\delta-1)(1-1/q)\rceil,\geq \delta]]_q$
stabilizer code that is pure to~$\delta$.
\end{theorem}
\begin{proof}
If we combine Lemmas~\ref{th:bchdim} and \ref{th:bcheuclideandual} and
apply the CSS construction, then we obtain the claim.
\end{proof}

One can argue in a similar way for hermitian duals of primitive,
narrow-sense BCH codes.
\begin{theorem}\label{th:bchhermitiandual} 
If $C$ is a narrow-sense primitive BCH code over $\F_{q^2}^n$ with
design distance $2\leq \delta \leq q^{m}-1$, then there exists an
$[[q^{2m}-1,q^{2m}-1-2m\lceil(\delta-1)(1-1/q^2)\rceil,\geq
\delta]]_q$ stabilizer code that is pure to $\delta$.
\end{theorem}
\begin{proof}
See~\cite{preprint0501126} for details. 
\end{proof}

When $m=1$, the BCH codes are the same as the Reed Solomon codes and
this case has been dealt with in \cite{grassl04}. An alternate perspective
using Reed-Muller codes is considered in \cite{klappenecker05p1}.

\paragraph{Extending quantum BCH codes.}
It is not always possible to extend a stabilizer code, because the
corresponding classical codes are required to be self-orthogonal.  In
this paragraph we will show that it is possible to extend 
narrow-sense BCH codes of certain lengths.

\begin{lemma}\label{th:bch_extension}
Let $\F_{q^2}$ be a finite field of characteristic $p$.  If $C$ is a
narrow-sense $[n,k,\geq d]_{q^2}$ BCH code such that $C^\hdual
\subseteq C$ and $n\equiv -1 \mod p$, then there exists an
$[[n,2k-n,\geq d]]_q$ stabilizer code that is pure to~$d$ which can be
extended to an $[[n+1,2k-n-1,\geq d+1]]_q$ stabilizer code that is
pure to~$d+1$.
\end{lemma}

\begin{proof}
Since $C^\hdual \subseteq C$, Corollary~\ref{co:classical} implies the
existence of an $[[n,2k-n,\geq d]]_q$ quantum code that is pure to~$d$. And
being narrow-sense the parity check matrix of $C$ has the form
\begin{eqnarray}
H&=&\left( \begin{array}{ccccc}
1 &\alpha &\alpha^{2}&\cdots&\alpha^{(n-1)}\\
1 &\alpha^{2} &\alpha^{2(2)}&\cdots&\alpha^{2(n-1)}\\
\vdots &\ddots &\ddots &\ddots &\ddots\\
1&\alpha^{d-1}&\alpha^{2(d-1)}&\cdots&\alpha^{(n-1)(d-1)}
\end{array}\right),
\end{eqnarray}
where $\alpha$ is a primitive $n^{th}$ root of unity. This can be
extended to give an $[n+1,k,d+1]$ code $C_e$, whose parity check
matrix is given as
\begin{eqnarray}
H_{e} &=&\left( \begin{array}{cccccc}
1 &1 &1&\cdots&1& 1\\
1 &\alpha &\alpha^{2}&\cdots&\alpha^{(n-1)}& 0\\
1 &\alpha^{2} &\alpha^{2(2)}&\cdots&\alpha^{2(n-1)}&0\\
\vdots &\ddots &\ddots &\ddots &\ddots&\vdots\\
1&\alpha^{d-1}&\alpha^{2(d-1)}&\cdots&\alpha^{(n-1)(d-1)}&0
\end{array}\right).
\end{eqnarray}
We will show that $C_e^\hdual$ is self-orthogonal. Let $R_i$ be the
$i^{th}$ row in $H_{e}$. For $2 \leq i\leq d$ the self-orthogonality
of $H$ implies that $\langle R_i|R_j\rangle_h=0$. We need to show
that $\langle R_i|\mbf{1}\rangle_h=0$, $1\leq i\leq d $. For $2\leq
i\leq d$ we have $\langle R_i|\mbf{1}\rangle_h
=\sum_{j=0}^{n-1}\alpha^{ij}=(\alpha^{in}-1)/(\alpha^{i}-1)=0$, as
$\alpha^n=1$ and $\alpha^i\neq 1$. For $i=1$ we have $\langle
\mbf{1}|\mbf{1}\rangle_h=n+1 \mod p$, which vanishes because of the
assumption $n\equiv -1 \mod p$.

Now we will show that the rank of $H_{e}$ is  $d$, thus $C_e$ has a
minimum distance of at least $d+1$. Any $d$ columns of $H_{e}$
excluding the last column form a $d\times d$ vandermonde matrix
which is nonsingular, indicating that the $d$ columns are linearly
independent. If we consider any set of $d$ columns that includes the
last column, we can find the determinant of the corresponding matrix
by expanding by the last column. This gives us a $d-1 \times d-1$
vandermonde matrix with nonzero determinant. Thus any $d$ columns of
$H_e$ are independent and the minimum distance of $C_e$ is at least
$d+1$. Therefore $C_e$ is an $[n+1,k,\geq d+1]_{q^2}$ extended cyclic code such that
$C_e^\hdual \subseteq C_e$. By Corollary~\ref{co:classical} it defines
an $[[n+1,2k-n-1,\geq d+1]]_q$ quantum code pure to $d+1$.
\end{proof}

\begin{corollary}
For all prime powers $q$, integers $m\ge 1$ and all $\delta$ in the range 
$2\leq \delta \leq q^{m}-1$ there exists an 
$$[[q^{2m},q^{2m}-2-2m\lceil(\delta-1)(1-1/q^2)\rceil,\geq \delta+1]]_q$$ 
stabilizer code pure to $\delta+1$. 
\end{corollary}
\begin{proof}
The stabilizer codes from Theorem~\ref{th:bchhermitiandual} are
derived from primitive, narrow-sense BCH codes. If $p$ denotes the
characteristic of $\F_{q^2}$, then $q^{2m}-1\equiv -1\bmod p$, so the
stabilizer codes given in Theorem~\ref{th:bchhermitiandual} can be
extended by Lemma~\ref{th:bch_extension}.  
\end{proof}

A result similar to Lemma~\ref{th:bch_extension} can be developed for
BCH codes that contain their euclidean duals.

\section{Puncturing Stabilizer Codes}
If we delete one coordinate in all codewords of a classical code, then
we obtain a shorter code that is called the punctured code. In
general, we cannot proceed in the same way with stabilizer codes,
since the resulting matrices might not commute if we delete one or
more tensor components. 

Rains~\cite{rains99} invented an interesting approach that solves the
puncturing problem for linear stabilizer codes and, even better, gives
a way to construct stabilizer codes from arbitrary linear codes. The
idea is to associate with a classical linear code a so-called puncture
code; if the puncture code contains a codeword of weight $r$, then a
self-orthogonal code of length $r$ exists and the minimum distance is
the same or higher than that of the initial classical code.  Further
convenient criteria for puncture codes were given in~\cite{grassl04}.

In this section, we generalize puncturing to arbitrary stabilizer
codes and review some known facts. Determining a puncture code is a
challenging task, and we conclude this section by showing how to
puncture quantum BCH codes.

\paragraph{The Puncture Code.} 
It will be convenient to denote the the pointwise product of two
vectors $u$ and $v$ in $\F_q^n$ by $uv$, that is, $uv=(u_iv_i)_{i=1}^n$.

Suppose that $C\le \F_q^{2n}$ is an arbitrary additive code. The
associated puncture code $\puncture_s(C) \subseteq \F_q^n$ is defined
as
$$ \puncture_s(C)= \left\{ (b_k a_k' - b_k'a_k)_{k=1}^n\,
| \,(a|b), (a'|b')\in C\right\}^\perp.$$

\begin{theorem} \label{th:punc_symplectic}
Suppose that $C$ is an arbitrary additive subcode of $\F_q^{2n}$ of
size $|C|=q^n/K$ such that $\swt(C^\sdual\setminus C)=d$.  If the
puncture code $\puncture_s(C)$ contains a codeword of Hamming weight
$r$, then there exists an $((r,K^*,d^*))_q$ stabilizer code with
$K^*\ge K/q^{n-r}$ that has minimum distance $d^* \ge d$ when $K^*>1$. If $\swt(C^\sdual)=d$, then the resulting
punctured stabilizer code is pure to $d$.
\end{theorem}
\begin{proof}
Let $x$ be a codeword of weight $r$ in the $\puncture_s(C)$. Define an
additive code $C_x\le \F_q^{2n}$ by
$$ C_x = \{ (a|bx) \; |\; (a|b)\in C\}.$$ 
If $(a|bx)$ and $(a'|b'x)$ are arbitrary elements of $C_x$, then 
\begin{equation}\label{puncturedual}
\< (a|bx)\; |\;(a'|b'x)>s
= \ds \tr\left(\sum_{k=1}^n (b_ka'_k - b_k'a_k)x_k\right)=0
\end{equation}
by definition of $\puncture_s(C)$; thus, $C_x\le (C_x)^\sdual.$ 

Let $C_x^R= \{ (a_k|b_k)_{k\in S} | (a|b)\in C_x\}$ denote the
restriction of $C_x$ to the support $S$ of the vector $x$. Since
equation (\ref{puncturedual}) depends only on the nonzero coefficients
of the vector $x$, it follows that $C_x^R \le (C_x^R)^{\sdual}$ holds.

We note that $|C|\ge |C_x^R|$; hence, the dimension $K^*$ of the
punctured quantum code is bounded by
\[
 K^* \geq q^r/|C_x^R| \geq q^r/|C|=q^r/(q^n/K)=K/q^{n-r}.
\]

It remains to show that $\swt((C_x^R)^\sdual\setminus C_x^R)\ge d$.
Seeking a contradiction, we suppose that $u_x^R$ is a vector in
$(C_x^R)^\sdual\setminus C_x^R$ such that $\swt(u_x^R)<d$. Let
$u_x=(a|b)$ denote the vector in $(C_x)^\sdual$ that is zero outside
the support of $x$ and coincides with $u_x^R$ when restricted to the
support of $x$. It follows that $(ax|b)$ is contained in $C^\sdual$.
However $\swt(ax|b)<d$, so $(ax|b)$ must be
an element of $C$, since $\swt(C^\sdual\setminus
C)=d$.  This implies that $(ax|bx)$ is an element of $C_x\le
(C_x)^\sdual$. Arguing as before, it follows that $(ax^2|bx)$ is in
$C$ and $(ax^2|bx^2)$ is in $C_x$.  Repeating the process, we obtain
that $v_x=(ax^{q-1}|bx^{q-1})$ is in $C_x$, and we note that $x^{q-1}$
is the characteristic vector of the support of $x$.  Restricting $v_x$
in $C_x$ to the support of $x$ yields $u_x^R\in C_x^R$,
contradicting the assumption that $u_x^R\in (C_x^R)^\sdual\setminus
C_x^R$.

Finally, the last statement concerning the purity is easy to prove (a
direct generalization of the argument given in~\cite{grassl04} for
pure linear codes).
\end{proof}

If the code $C$ is a direct product, as in the case of CSS codes, then 
the expression for the puncture code simplifies somewhat. 
\begin{lemma}
If $C_1$ and $C_2$ are two additive subcodes of $\F_q^n$, then 
$$ \pc_s(C_1\times C_2) = \{ ab\mid a\in C_1, b\in
C_2\}^\perp\le \F_q^n.$$
\end{lemma}
\begin{proof}
Since $\langle ab\mid a\in C_1, b\in C_2\rangle = \langle
(ba'-b'a)\mid a,a'\in C_1, b,b'\in C_2\rangle$, the claim about the
orthogonal complements of these sets is obvious.
\end{proof}

Since many quantum codes are constructed from self-orthogonal codes $C
\le C^\perp$, we write shortly
\begin{equation}
\pc_e(C) =\pc_s(C\times C)= \{ ab \mid a,b\in C\}^\perp.
\end{equation}

\paragraph{Puncturing BCH Codes.} 
In this section, we let $\B_q^m(\delta)$ denote a primitive,
narrow-sense $q$-ary BCH code of length $n=q^m-1$ and designed
distance $\delta$. We will illustrate the previous result by
puncturing such BCH codes. Some knowledge about the puncture code is
necessary for this task, and we show in Theorem~\ref{th:bchP(C)} that
a cyclic generalized Reed-Muller code is contained in the puncture
code.

First, let us recall some basic facts about cyclic generalized Reed-Muller
codes, see~\cite{assmus92,assmus98,kasami68,pellikaan04} for
details. Let $L_m(\nu)$ denote the subspace of $\F_q[x_1,\dots,x_m]$
of polynomials of degree~$\le \nu$, and let $(P_0,\dots, P_{n-1})$ be
an enumeration of the points in $\F_q^m$ where $P_0=\mbf{0}$.  The $q$-ary cyclic
generalized Reed-Muller code $\RM^*_q(\nu,m)$ of order $\nu$ and
length $n=q^m-1$ is defined as
$$ \RM^*_q(\nu,m) = \{ ev\, f\,|\, f\in
L_m(\nu)\},$$ where the codewords are evaluations of the polynomials in
all but $P_0$, $ev\, f=(f(P_1),\dots, f(P_{n-1}))$. 
The dimension $k^*(\nu)$ of the code 
$\RM^*_q(\nu,m)$ is given by the formula
$
k^*(\nu) = \sum_{j=0}^m (-1)^j \binom{m}{j}\binom{m+\nu-jq}{\nu-jq}, 
$
and its minimum distance $d^*(\nu)=(R+1)q^Q-1,$
where $m(q-1)-\nu = (q-1)Q+ R$ with $0\le R<q-1$.
The dual code of $\RM_q^*(\nu,m)$ can be characterized by  
\begin{eqnarray}
\RM_q^*(\nu,m)^\perp
&=&\{ ev \mbox{ }f \,|\, f\in L_m^*(\nu^\perp)
\},\label{eq:RMdualdefn}
\end{eqnarray}
where $\nu^\perp= m(q-1)-\nu-1$ and $L_m^*(\nu)$ is the subspace of
all nonconstant polynomials in $L_m(\nu)$; 

It is well-known that a primitive, narrow-sense BCH code contains a
cyclic generalized Reed-Muller code, see~\cite[Theorem 5]{kasami68},
and we determine the largest such subcode in our next lemma.

\begin{lemma}\label{th:largestRMinbch}
We have $\RM_q^*(\nu,m)\subseteq \B_q^m(\delta)$ for $\nu = (m-Q)(q-1)-R$,
with $Q=\lfloor\log_q(\delta+1) \rfloor$ and $R=\lceil
(\delta+1)/q^Q\rceil-1$. For all orders $\nu'>\nu$, we have 
$\RM_q^*(\nu',m)\not\subseteq \B_q^m(\delta)$.
\end{lemma}

\begin{proof}
First, we show that $\RM_q^*(\nu,m) \subseteq \B_q^m(\delta)$.  Recall
that the minimum distance $d^*(\nu)=(R+1)q^Q-1$, where
$m(q-1)-\nu=(q-1)Q+R$ with $0\le R<q-1$.  By \cite[Theorem
5]{kasami68}, we have $\RM_q^*(\nu,m) \subseteq \B_q^m((R+1)q^Q-1)$.
Notice that $(R+1)q^Q-1 =\lceil (\delta+1)/ q^Q\rceil q^Q -1 \geq
\delta$, so $\B_q^m((R+1)q^Q-1) \subseteq \B_q^m(\delta)$. Therefore,
$\RM_q^*(\nu,m) \subseteq \B_q^m(\delta)$, as claimed.

For the second claim, it suffices to show that $\RM_q^*(\nu+1,m)$ is
not a subcode of $\B_q^m(\delta)$. We will prove this by showing that
the minimum distance $d^*(\nu+1)< \delta$.  Notice that 
\begin{eqnarray*}
m(q-1)-(\nu+1) 
&=&\left\{\begin{array}{ll} (q-1)Q+R-1 
& \text{for } R\geq 1,\\
(q-1)(Q-1)+q-2& \text{for } R=0. \end{array} \right.
\end{eqnarray*}
with $R$ and $Q$ as given in the hypothesis. 
Therefore, the distance $d^*(\nu+1)$ of 
$\RM_q^*(\nu+1,m)$ is given by 
\begin{eqnarray*}
d^*(\nu+1) &=& 
\left\{\begin{array}{ll} (\lceil (\delta+1)/q^Q\rceil-1)q^Q-1 
&\text{for } R\geq 1,\\
(q-1)q^{Q-1}-1&\text{for } R=0. \end{array} \right.
\end{eqnarray*}
In both cases, it is straightforward to verify that $d^*(\nu+1)<\delta$. 
\end{proof}

Explicitly determining the puncture code is a challenging task. For
the duals of BCH codes, we are able to determine large subcodes of the
puncture code.

\begin{theorem}\label{th:bchP(C)}
If $ \delta < q^{\lfloor m/2 \rfloor}-1$, then $\RM_q^*(\mu,m)
\subseteq \pc_e(\B_q^m(\delta)^\perp)$ for all orders $\mu$ in the range
$0\leq \mu\leq m(q-1)-2(R+(q-1)Q)+1$ with $Q=\lfloor\log_q
(\delta+1)\rfloor$ and $R=\lceil(\delta+1)/q^Q\rceil -1$.
\end{theorem}
\begin{proof}
By Lemma~\ref{th:largestRMinbch}, we have $\RM_q^*(\nu,m)
\subseteq \B_q^m(\delta)$ for $\nu= (m-Q)(q-1)-R$; hence,
$\B_q^m(\delta)^\perp \subseteq \RM_q^*(\nu,m)^\perp$.
It follows from the 
definition of the puncture code that 
$\pc_e(\B_q^m(\delta)^\perp) \supseteq \pc_e(\RM_q^*(\nu,m)^\perp)$. However, 
\begin{eqnarray*}
\pc_e(\RM_q^*(\nu,m)^\perp) &=&\{ev f \cdot ev \mbox{ } g \mid f,g \in
L_m^*(\nu^\perp)\}^\perp,\\ &\supseteq & \{ ev f \mid f \in
L_m^*(2\nu^\perp)\}^\perp=\RM_q^*((2\nu^\perp)^\perp,m),
\end{eqnarray*}
where the last equality follows from equation~(\ref{eq:RMdualdefn}).  This is
meaningful only if $(2\nu^\perp)^\perp \geq 0$ or, equivalently, if 
$\nu\geq(m(q-1)-1)/2$. Since $\delta < q^{\lfloor m/2\rfloor}-1$, it follows that $Q\leq \lfloor m/2\rfloor -1$, and the order $\nu$ satisfies
$$\begin{array}{lcl}
\nu&=&(m-Q)(q-1)-R \geq \lceil m/2+1\rceil(q-1)-R \\
&\geq& \lceil m/2\rceil (q-1)+1 \geq (m(q-1)-1)/2,
\end{array}
$$
as required. Since $\RM_q^*(\mu,m) \subseteq \RM_q^*((2\nu^\perp)^\perp,m)$ for 
$0\leq \mu\leq (2\nu^\perp)^\perp$, we have $\RM_q^*(\mu,m) \subseteq \pc_e(\B_q^m(\delta)^\perp)$.
\end{proof}

Unfortunately, the weight distribution of generalized cyclic
Reed-Muller codes is not known, see~\cite{charpin98}.  However, we
know that the puncture code of $\B_q^m(\delta)^\perp$ contains the
codes $\RM_q^*(0,m)\subseteq \RM_q^*(1,m)\subseteq \cdots \subseteq
\RM_q^*(m(q-1)-2(R+(q-1)Q)+1,m)$, so it must contain codewords of the
respective minimum distances.

\begin{corollary} \label{th:bchpunc}
If $\delta$ and $\mu$ are integers in the range $2\leq \delta <
q^{\lfloor m/2\rfloor}-1$ and $0\leq \mu\leq m(q-1)-2(R+(q-1)Q)+1$,
where $Q=\lfloor\log_q(\delta+1)\rfloor$ and
$R=\lceil(\delta+1)/q^Q\rceil-1$, then there exists
a 
$$[[d^*(\mu), \geq d^*(\mu)-2m\lceil(\delta-1)(1-1/q)\rceil,\geq
\delta]]_q$$ stabilizer code of length $d^*(\mu)=(\rho+1)q^\sigma-1$, 
where $\sigma$ and $\rho$ satisfy 
the relations $m(q-1)-\mu=(q-1)\sigma+\rho$ and $0\leq \rho<q-1$.
\end{corollary}

\begin{proof}
If $2\leq \delta < q^{\lfloor m/2\rfloor }-1 $, then from Theorem~\ref{co:bcheuclideandual} 
we know that there exists an
$[[q^m-1,q^m-1-2m\lceil(\delta-1)(1-1/q)\rceil,\geq \delta]]_q$ quantum code.
From Lemma~\ref{th:bchP(C)} we know that
$\pc_e(\B_q^m(\delta)^\perp)\supseteq \RM_q^*(\mu,m)$, where $0\leq
\mu\leq m(q-1)-2(q-1)Q-2R+1$.  By Theorem~\ref{th:punc_symplectic}, if
there exists a vector of weight $r$ in $\pc_e(\B_q^m(\delta)^\perp)$,
the corresponding quantum code can be punctured to give $[[r,\geq
r- 2m\lceil(\delta-1)(1-1/q)\rceil) ,d\geq \delta]]_q$.  The minimum distance
of $\RM_q^*(\mu,m)$ is $d^*(\mu)=(\rho+1)q^\sigma-1$, where 
$0\leq \rho<q-1$ \cite[Theorem 5]{kasami68}. Hence, it is always possible to puncture
the quantum code to $[[d^*(\mu), \geq d^*(\mu)-2m\lceil(\delta-1)(1-1/q)\rceil,
\geq \delta ]]_q$.
\end{proof}

It is also possible to puncture quantum codes constructed via
classical codes self-orthogonal with respect to the hermitian inner
product. Examples of such puncturing can be found in \cite{grassl04}
and \cite{klappenecker05p1}.

\section{MDS Codes}\label{sec:MDS} 
A quantum code that attains the quantum Singleton bound is called a
quantum Maximum Distance Separable code or quantum MDS code for short.
These codes have received much attention, but many aspects have not
yet been explored in the quantum case (but
see~\cite{grassl04,rains99}).  In this section we will study the
maximal length of MDS stabilizer codes.

An interesting result concerning the purity of quantum MDS codes was
derived by Rains~\cite[Theorem 2]{rains99}: 

\begin{lemma}[Rains] \label{th:d_purity}
An $[[n,k,d]]_q$ quantum MDS code with $k\ge 1$ is pure up to $n-d+2$.
\end{lemma}

\begin{corollary} \label{th:mds_purity}
All quantum MDS codes are pure.
\end{corollary}
\begin{proof}
An $[[n,k,d]]_q$ quantum MDS code with $k=0$ is pure by definition; if
$k\ge 1$ then it is pure up to $n-d+2$.  By the quantum Singleton
bound $n-2d+2=k\geq 0$; thus, $n-d+2\ge d$, which means that the code
is pure.
\end{proof}

\begin{lemma} \label{th:mds_classical}
For any $[[n,n-2d+2,d]]_q$  quantum MDS stabilizer code with $n-2d+2>0$ the
corresponding classical codes $C\subseteq C^\adual$ are also MDS. 
\end{lemma}
\begin{proof}
If an $[[n,n-2d+2,d]]_q$ stabilizer code exists, then
Theorem~\ref{th:alternating} implies the existence of an additive
$[n,d-1]_{q^2}$ code $C$ such that $C\subseteq C^\adual$.
Corollary~\ref{th:mds_purity} shows that $C^\adual$ has minimum
distance $d$, so $C^\adual$ is an $[n,n-d+1,d]_{q^2}$ MDS code. By
Lemma~\ref{th:d_purity}, the minimum distance of $C$ is $\ge n-d+2$,
so $C$ is an $[n,d-1,n-d+2]_{q^2}$ MDS code.
\end{proof}

A classical $[n,k,d]_q$ MDS code is said to be trivial if $k\leq 1 $
or $k\geq n-1$. A trivial MDS code can have arbitrary length, but a
nontrivial one cannot. The next lemma is a straightforward
generalization from linear to additive MDS codes.

\begin{lemma} \label{th:mds_nontrivial}
Assume that there exists a classical additive $(n,q^k,d)_q$ MDS code $C$.
\begin{compactenum}[(i)]
\item If the code is trivial, then it can have arbitrary length.
\item If the code is nontrivial, then its code parameters must be in the range 
$2\leq k\leq \min\{ n-2,q-1\}$ and
$n\leq q+k-1\leq 2q-2$.
\end{compactenum}
\end{lemma}
\begin{proof} The first statement is obvious. For (ii), we note that 
the weight distribution of the code $C$ and its dual are related by the
MacWilliams relations. The proof given in
\cite[p.~320-321]{macwilliams77} for linear codes applies without
change, and one finds that the number of codewords of weight $n-k+2$
in $C$ is given by
$$ A_{n-k+2}=\binom{n}{k-2}(q-1)(q-n+k-1).$$
Since $A_{n-k+2}$ must be a nonnegative number, we obtain the claim. 
\end{proof}

We say that a quantum $[[n,k,d]]_q$ MDS code is trivial if and only if
its minimum distance $d\le 2$.  The length of trivial quantum MDS
codes is not bounded, but the length of nontrivial ones is, 
as the next lemma shows.

\begin{theorem}[Maximal Length of MDS Stabilizer Codes]\label{th:mds_length}
A nontrivial $[[n,k,d]]_q$  MDS stabilizer code satisfies 
the following constraints: 
\begin{compactenum}[i)]
\item its length $n$ is in the range $4\leq n \leq q^2+d-2\leq 2q^2-2$;
\item its minimum distance satisfies $\max\{3,n-q^2+2\} 
\leq d \leq \min\{ n-1,q^2\}$. 
\end{compactenum}
\end{theorem}
\begin{proof}
By definition, a quantum MDS code attains the Singleton bound, so
$n-2d+2=k\ge 0$; hence, $n\ge 2d-2$. Therefore, a nontrivial quantum
MDS code satisfies $n\ge 2d-2\ge 4$.

By Lemma \ref{th:mds_classical}, the existence of an
$[[n,n-2d+2,d]]_q$  stabilizer code implies the existence of
classical MDS codes $C$ and $C^\adual$ with parameters
$[n,d-1,n-d+2]_{q^2}$ and $[n,n-d+1,d]_{q^2}$, respectively. If the
quantum code is a nontrivial MDS code, then the associated classical
codes are nontrivial classical MDS codes. Indeed, for $n\ge 4$ the quantum
Singleton bound implies $d\le (n+2)/2\le (2n-2)/2=n-1$, so $C$ is a
nontrivial classical MDS code. 

By Lemma \ref{th:mds_nontrivial}, the dimension of $C$ satisfies the
constraints $2\leq d-1 \leq \min\{n-2,q^2-1\}$, or equivalently $3\leq d\leq
\min\{n-1,q^2 \}$. Similarly, the length $n$ of $C$ satisfies 
$n\leq q^2+(d-1)-1 \leq 2q^2-2$. If we combine these inequalities then
we get our claim.
\end{proof}

\begin{example} 
The length of a nontrivial binary MDS stabilizer code cannot exceed
$2q^2-2=6$. In \cite{calderbank98} the nontrivial MDS stabilizer codes
for $q=2$ were found to be $[[5,1,3]]_2$ and $[[6,0,4]]_2$, so there
cannot exist further nontrivial MDS stabilizer codes.
\end{example}

In~\cite{grassl04}, the question of the maximal length of MDS codes
was raised. All MDS stabilizer codes provided in that reference had a
length of $q^2$ or less; this prompted us to look at the following
famous conjecture for classical codes (cf.~\cite[Theorem~7.4.5]{huffman03} or
\cite[pages~327-328]{macwilliams77}).

\smallskip

\noindent\textbf{MDS Conjecture.} \textit{
If there is a nontrivial $[n,k]_q$ MDS code, then $n\leq q+1$ except
when $q$ is even and $k=3$ or $k=q-1$ in which case $n\leq
q+2$.}
\smallskip

If the MDS conjecture is true (and much supporting evidence is
known), then we can improve upon the result of
Theorem~\ref{th:mds_length}.

\begin{corollary}
If the classical MDS conjecture holds, then there are no nontrivial
MDS stabilizer codes of lengths exceeding $q^2+1$ except when
$q$ is even and $d=4$ or $d=q^2$ in which case $n\leq q^2+2$.
\end{corollary}

\section{Quantum Character Codes} 
A new family of codes was introduced in~\cite{ding00}.  The codes of
this family are defined using group characters. These codes are in
many ways remarkably similar to binary Reed-Muller codes, but they are
defined over nonbinary fields.  Since these codes were introduced only
recently and are not yet well-known, we will provide a little more
background. In this section we derive quantum
codes from group character codes using the CSS construction.  

\paragraph{Group character codes.}
Let $G$ be an additive abelian group of order $n$ and exponent $m$.
Let $\F_q$ be a finite field such that $\gcd(n,q)=1$ and $m\mid
q-1$. 

The set $\Hom(G,\F_q^*)$ of $\F_q$-valued characters of $G$ consists
of the homomorphisms from $G$ into the multiplicative group
$\F_q^*$. Our assumptions ensure that the set of characters forms a
group that is isomorphic to~$G$. We can index the characters by
elements of the group $G$,
$$\Hom(G,\F_q^*)=\{ \chi_x\,|\, x\in G\},$$ 
such that $\chi_0$ denotes the trivial character,
and $\chi_{-x}$ denotes the inverse of~$\chi_x$.

For any subset $X$ of the group $G$, 
the character code $C_X$ is defined as 
\begin{eqnarray}
C_X&=&\left\{c\in\F_{q}^n \Bigg|\sum_{i=0}^{n-1}c_i\chi_{x_i}(y)=0
\mbox{ for all }  y \in X\right \}. 
\end{eqnarray}
The code $C_X$ is an $[n,k]_q$ code with $n=|G|$ and $k=n-|X|$.  The
parity check matrix $H_X$ of the code $C_X$, with $X=\{x_0,\dots,
x_{n-k+1}\}$, is given by
$$ H_X=\left(\begin{array}{cccc}
\chi_{x_0}(x_0)&\chi_{x_1}(x_0)&\cdots&\chi_{x_{n-1}}(x_0)\\
\chi_{x_0}(x_1)&\chi_{x_1}(x_1)&\cdots&\chi_{x_{n-1}}(x_1)\\
\vdots&\vdots &\ddots &\vdots\\
\chi_{x_0}(x_{n-k-1})&\chi_{x_1}(x_{n-k-1})&\cdots&\chi_{x_{n-1}}(x_{n-k-1})
\end{array}\right),
$$
and its generator matrix $G_X$ by
\begin{equation}\label{eq:def_grp2}
G_X=\left(\begin{array}{cccc}
\chi_{x_0}(-x_{n-k})&\chi_{x_1}(-x_{n-k})&\cdots&\chi_{x_{n-1}}(-x_{n-k})\\
\chi_{x_0}(-x_{n-k+1})&\chi_{x_1}(-x_{n-k+1})&\cdots&\chi_{x_{n-1}}(-x_{n-k+1})\\
\vdots&\vdots &\ddots &\vdots\\
\chi_{x_0}(-x_{n-1})&\chi_{x_1}(-x_{n-1})&\cdots&\chi_{x_{n-1}}(-x_{n-1})
\end{array}\right).
\end{equation}
Indeed, the characters satisfy the well-known orthogonality relation
\begin{eqnarray*}
\sum_{x\in G}\chi_x(y)\chi_x(z)
&=&\left\{\begin{array}{rl}n&\mbox{if }y+z=\mbf{0},\\0&\mbox{if
}y+z\neq \mbf{0};\end{array}\right. 
\end{eqnarray*}
which implies $G_XH^T_X=0$.

\paragraph{Elementary abelian 2-groups.} 
We now specialize to the case of a finite elementary abelian 2-group
$G=\mbf{Z}_2^m$, $m\ge 1$. Let $\F_q$ be a finite field of odd
characteristic; this choice ensures that $2 \mid q-1$ and
$\gcd(2^m,q)=1$. Recall that the characters of $G$ are given by
$\chi_x(y)=(-1)^{x\cdot y}$ for $x, y$ in $G$.

We define a 2-group character code $\mathcal{C}_{q}(r,m)$ by 
$$ \mathcal{C}_{q}(r,m) = C_X \quad \mbox{with} \quad 
X=\{x\in \mbf{Z}_2^m \mid \mbox{wt}(x) > r\}.$$
It can be shown
that $C_{q}(r,m)$ is an $[n,k(r),d(r)]_q$ code, with 
\begin{equation}\label{eq:grp2_dim}
k(r)=\sum_{j=0}^{r}\binom{m}{j} \quad\mbox{and}\quad
d(r)=2^{m-r},
\end{equation}
see~\cite[Lemma 4 and Theorem 6]{ding00}. We need the following result
about 2-group character codes which is not explicitly proved in
\cite{ding00}.
\begin{lemma} \label{th:grp2_contain}
If $r_1\leq r_2$, then $\mathcal{C}_q(r_1,m) \subseteq
\mathcal{C}_q(r_2,m)$.
\end{lemma}
\begin{proof}
By equation (\ref{eq:def_grp2}) the generator matrix of
$\mathcal{C}_q(r,m)$ consists of vectors of the form
$$
(\chi_{x_0}(x_{i}),\chi_{x_1}(x_{i}),\cdots,\chi_{x_{n-1}}(x_{i})) = 
(\chi_{x_0}(-x_{i}),\chi_{x_1}(-x_{i}),\cdots,\chi_{x_{n-1}}(-x_{i}))$$
where $x_i$ is an element of $\mbf{Z}_2^m$ of Hamming 
weight $\wt(x_i) \leq r$. Thus, the generator matrix of 
$C_q(r_1,m)$ is a submatrix of the generator matrix of 
$C_q(r_2,m)$, which shows that  
$\mathcal{C}_q(r_1,m)\subseteq \mathcal{C}_q(r_2,m)$.
\end{proof}

\begin{lemma} \label{th:grp2_dual}
The dual code $\mathcal{C}_q(r,m)^\perp$ is equivalent to
$\mathcal{C}_q(m-r-1,m)$.
\end{lemma}
\begin{proof}
See \cite[Theorem 8]{ding00}.
\end{proof}
Now we will construct a family of codes based on the CSS
construction.
\begin{theorem}\label{th:grp2_css0}
If $0\leq r_1< r_2\leq m$ and $q$ the power of an odd prime, then
there exists an $[[n, k(r_2)-k(r_1), \min\{2^{m-r_2},2^{r_1+1}\}
]]_{q}$ quantum code, where $n=2^m$ and $k(r)$ is given by equation
(\ref{eq:grp2_dim}).
\end{theorem}
\begin{proof}

If $r_1< r_2$, then $C_1=\mathcal{C}_{q}(r_1,m)\subseteq
\mathcal{C}_{q}(r_2,m)=C_2$ by Lemma \ref{th:grp2_contain}.  From the
equations for the minimum distances given in~(\ref{eq:grp2_dim}), we
can see that $\wt(C_2\setminus C_1) = 2^{m-r_2}$. Similarly, it follows
from Lemma~\ref{th:grp2_dual} that $\wt(C_1^\perp\setminus C_2^\perp)
=\wt(\mathcal{C}_q(m-r_1-1)\setminus
\mathcal{C}_q(m-r_2-1))=2^{r_1+1}$. By Lemma \ref{th:css}, there exists an
$[[n,k(r_2)-k(r_1), \min\{ 2^{m-r_2},2^{r_1+1}\}]]_{q}$ stabilizer code,
where the dimensions $k(r_1)$ and $k(r_2)$ are given by equation
(\ref{eq:grp2_dim}).
\end{proof}

We can get more quantum codes by puncturing, as we did in the case of
BCH codes. However, only the weight
distribution of $\mathcal{C}_q(1,m)$ is known, so at the moment we
do not have enough information as to what codes might exist.

\section{Code Constructions}
Constructing good quantum codes is a difficult task. We need a quantum
code for each parameter $n$ and $k$ in our tables. We collect in this
section some simple facts about the construction of
codes. Lemmas~\ref{th:lengthening}--\ref{th:smallerdim} show how to
lengthen, shorten or reduce the dimension of the stabilizer code.

\begin{table}[htb]
\begin{center}
\begin{tabular}{c||c|c|c}
n/k & $k-1$ & $k$ & $k+1$ \\
\hline\hline
$n-1$& \stacked{$d-1$ pure}{Lemma~\ref{th:smallerdim}}   &  \stacked{$d-1$ pure}{Lemma~\ref{th:smallerdim}}         & \stacked{$d-1$ pure}{Lemma~\ref{th:shorterlength}}     \\
\hline
$n$  & \stacked{$d$ pure}{Lemma~\ref{th:smallerdim}} & \fbox{$d$ pure}  &      \\
\hline
$n+1$& \stacked{$d$ impure}{Lemma~\ref{th:lengthening}} & \stacked{$d$ impure}{Lemma~\ref{th:lengthening}}&
\end{tabular}
\end{center}
\caption{The existence of a pure $[[n,k,d]]_q$ stabilizer code implies the existence of codes with other parameters.}\label{table:cc}
\end{table}

\begin{lemma}\label{th:lengthening}
If an $[[n,k,d]]_q$ stabilizer code exists for $k>0$, then there
exists an impure $[[n+1,k,d]]_q$ stabilizer code.
\end{lemma}
\begin{proof}
If an $[[n,k,d]]_q$ stabilizer code exists, then there exists an
additive subcode $C\le \F_q^{2n}$ such that $|C|=q^{n-k}$, $C\le
C^\sdual$, and $\swt(C^\sdual\setminus C)=d$. Define the additive code
$$ C' = \{ (a\alpha|b0)\,|\, \alpha\in \F_q, (a|b)\in \F_q^{2n}\}.$$
We have $|C'|=q^{n-k+1}$. The definition ensures that $C'$ is
self-orthogonal with respect to the trace-symplectic inner
product. Indeed, two arbitrary elements $(a\alpha|b0)$ and
$(a'\alpha'|b'0)$ of $C'$ satisfy the orthogonality condition
$$ \< (a\alpha|b0) | (a'\alpha'|b'0)>s = \< (a|b) | (a'|b')>s +
\tr(\alpha\cdot 0 - \alpha'\cdot 0) = 0.$$ A vector in the
trace-symplectic dual of $C'$ has to be of the form $(a\alpha|b0)$ with
$(a|b)\in C^\sdual$ and $\alpha\in \F_q$.
Furthermore,
$$ \swt(C'^\sdual \setminus C') = \min\{ \swt(a\alpha|b0)\,|\,
\alpha\in \F_q, a,b\in C^\sdual\setminus C\},$$ which coincides with
$\swt(C^\sdual\setminus C).$ Therefore, an $[[n+1,k,d]]_q$ stabilizer
code exists by Theorem~\ref{th:stabilizer}.  If $d>1$, then the code
is impure, because $C'^\sdual$ contains the vector
$(0\alpha|00)$ of symplectic weight 1.
\end{proof}

\begin{lemma}\label{th:shorterlength}
If a pure $[[n,k,d]]_q$ stabilizer code exists with $n\ge 2$ and $d\ge
2$, then there exists a pure $[[n-1,k+1,d-1]]_q$ stabilizer code.
\end{lemma}
\begin{proof}
If a pure $[[n,k,d]]_q$ stabilizer code exists, then there exists an
additive code $D\le \F_{q^2}^n$ that is self-orthogonal with respect
to the trace-alternating form, so that $|D|=q^{n-k}$ and
$\wt(D^\adual)=d$.  Let $D_0^\adual$ denote the code obtained by
puncturing the first coordinate of $D^\adual$. Since the minimum
distance of $D^\adual$ is at least 2, we know that
$|D_0^\adual|=|D^\adual|=q^{n+k}$, and we note that the minimum
distance of $D_0^\adual$ is $d-1$. The dual of $D_0^\adual$ consists
of all vectors $u$ in $\F_{q^2}^{n-1}$ such that $0u$ is contained
in~$D$. Furthermore, if $u$ is an element of $D_0$, then $0u$ is
contained in $D$; hence, $D_0$ is a self-orthogonal additive code.  The
code $D_0$ is of size $q^{(n-1)-(k+1)}$, because
$$\dim D_0+ \dim D_0^\adual = \dim \F_{q^2}^{n-1}$$
when we view $D_0$ and its dual as $\F_p$--vector spaces.
It follows that there exists a pure $[[n-1,k+1,d-1]]_q$ stabilizer code.
\end{proof}

\begin{lemma}\label{th:smallerdim}
If a (pure) $[[n,k,d]]_q$ stabilizer code exists, with $k\ge 2$ ($k\ge
1$), then there exists a (pure)\/ $[[n, k-1,d^*]]_q$ stabilizer code such that $d^*\ge d$.
\end{lemma}
\begin{proof}
If an $[[n,k,d]]_q$ stabilizer code exists, then there exists an
additive code $D\le \F_{q^2}^n$ such that $D\le D^\adual$ with
$\wt(D^\adual\setminus D)=d$ and $|D|=q^{n-k}$. Choose an additive
code~$D_b$ of size $|D_b|=q^{n-k+1}$ such that $D\le D_b\le D^\adual$.
Since $D\le D_b$, we have $D_b^\adual\le D^\adual$.  The set $\Sigma_b
= D_b^\adual \setminus D_b$ is a subset of $D^\adual \setminus D$,
hence the minimum weight $d^*$ of $\Sigma_b$ is at least $d$.
This proves the existence of an $[[n,k-1,d^*]]$ code.

If the code is pure, then $\wt(D^\adual)=d$; it follows from
$D_b^\adual \le D^\adual$ that $\wt(D_b^\adual)\ge d$, so the smaller
code is pure as well.
\end{proof}

\begin{corollary}
If a pure $[[n,k,d]]_q$ stabilizer code with $n\ge 2$ and
$d\ge 2$ exists, then there exists a pure $[[n-1,k,d-1]]_q$ stabilizer
code.
\end{corollary}
\begin{proof}
Combine Lemmas~\ref{th:shorterlength} and \ref{th:smallerdim}.
\end{proof}

\begin{lemma}\label{th:directsum}
Suppose that an $((n,K,d))_q$ and an $((n',K',d'))_q$ stabilizer code
exists.  Then there exists an $((n+n', KK',\min(d,d'))_q$ stabilizer
code.
\end{lemma}
\begin{proof}
Suppose that $P$ and $P'$ are the orthogonal projectors onto the
stabilizer codes for the $((n,K,d))_q$ and $((n',K',d'))_q$ stabilizer
codes, respectively. Then $P\otimes P'$ is an orthogonal projector
onto a $KK'$-dimensional subspace $Q^*$ of $\C^d$, where $d=q^{n+n'}$.
Let $S$ and $S'$ respectively denote the stabilizer groups of the
images of $P$ and $P'$. Then $S^*=\{ E\otimes E'\,|\, E\in S, E'\in
S'\}$ is the stabilizer group of $Q^*$.

If an element $F\otimes F^*$ of $G_n\otimes G_{n'}=G_{n+n'}$ is not
detectable, then $F$ has to commute with all elements in $S$, and $F'$
has to commute with all elements in $S'$.  It is not possible that both
$F\in Z(G_n)S$ and $F'\in Z(G_{n'})S'$ hold, because this would imply that
$F\otimes F'$ is detectable. Therefore, either $F$ or $F'$ is not
detectable, which shows that the weight of $F\otimes F'$ is at least
$\min(d,d')$.
\end{proof}

\begin{lemma}
Let $Q_1$ and $Q_2$ be pure stabilizer codes that respectively have parameters
$[[n,k_1,d_1]]_q$ and $[[n,k_2,d_2]]$. If\/
$Q_2\subseteq Q_1$, then there exists a $[[2n,k_1+k_2,d]]_q$
pure stabilizer code with minimum distance $d\ge \min\{ 2d_2,d_1\}$.
\end{lemma}
\begin{proof}
The hypothesis implies that there exist
additive subcodes $D_1\le D_2$ of\/ $\F_{q^2}^n$ such that
 $D_m\le D_m^\adual$, $|D_m|=q^{n-k_m}$, and
$\wt(D_m^\adual)=d_m$ for $m=1,2$.  The additive code
$$ D = \{ (u,u+v)\;|\, u\in D_1, v\in D_2\}\le \F_{q^2}^{2n}$$ is of
size $|D|=q^{2n-(k_1+k_2)}$. The trace-alternating dual of the code
$D$ is $D^\adual = \{ (u'+v',v')\,|\, u'\in D_1^\adual, v'\in
D_2^\adual\}$.  Indeed, the vectors on the right hand side are
perpendicular to the vectors in $D$, because
$$ \( (u,u+v)\, |\, (u'+v',v') )a =
\( u| u'+v')a + \( u+v|v')a = 0
$$
holds for all $u\in D_1, v\in D_2$ and $u'\in D_1^\adual, v'\in
D_2^\adual$.  We observe that $D$ is self-orthogonal, $D\le D^\adual$.
The weight of a vector $(u'+v',v')\in D^\adual\setminus D$ is at least
$\min\{ 2d_2,d_1\}$; the claim follows.
\end{proof}

\begin{lemma}
\label{th:difference}
Let $q$ be an even prime power.  If a pure $[[n,k_1,d_1]]_q$ stabilizer
code~$Q_1$ exists that has a pure subcode $Q_2\subseteq Q_1$ with
parameters $[[n,k_2,d_2]]_q$ such that $k_1>k_2$, then a
pure $[[2n,k_1-k_2,d]]_q$ stabilizer code exists such that $d \ge
\min{\{2d_1, d_2\}}$.
\end{lemma}
\begin{proof}
If an $[[n_m,k_m,d_m]]_q$ stabilizer code exists, then there exists an
additive code $D_m\le \F_{q^2}^n$ such that $D_m\le D_m^\adual$,
$\wt(D_m^\adual)=d$, and $|D_m|=q^{n-k_m}$ for $m=1,2$.
The inclusion $Q_2\subseteq Q_1$ implies that $D_1\le D_2$.
Let $D$ denote the additive code consisting of vectors of the form
$(u,u+v)$ such that $u \in D_2^\adual$ and $v \in D_1$.

We claim that $D^\adual$ consists of vectors of the form $(u',u'+v')$
such that $u' \in D_1^\adual$ and $v' \in D_2$.  Indeed, let
$v_1=(u,u+v)$ denote a vector in~$D$, and let $v_2 = (u',u'+v')$ be a
vector with $u' \in D_1^\adual$ and $v' \in D_2$.  We have
$$\(v_1|v_2)a= \(u|u')a + \(u|u')a + \(u|v')a + \(v|u')a + \(v|v')a.$$
The first two terms on the right hand side cancel because the
characteristic of the field is even; the next two terms vanish since
the vectors belong to dual spaces; the last term vanishes because $v$
and $v'$ are both contained in $D_2$, and $D_2$ is self-orthogonal.
Therefore, $v_1$ and $v_2$ are orthogonal. The set $\{ (u',u'+v')\,|\,
u' \in D_1^\adual, v' \in D_2\}\subseteq D^\adual$ has cardinality
$q^{2n+k_1-k_2}$, so it must be equal to $D^\adual$ by a dimension
argument.

The Hamming weight of a vector $(u',u'+v')$ in $D^\adual$
is at least $\min{\{2d_1, d_2\}}$, because
$u' \in D_1^\adual$ and $v'\in D_2\le D_2^\adual$.
\end{proof}

\begin{lemma} \label{th:codeexpansion} 
Let $q$ be a power of a prime.  If an $((n,K,d))_{q^m}$ stabilizer
code exists, then an $((nm,K,\ge d))_q$ stabilizer code
exists. Conversely, if an $((nm,K,d))_q$ stabilizer code exists,
then there exists an $((n,K,\ge \lfloor d/m\rfloor))_{q^m}$ stabilizer code.
\end{lemma}
This lemma is implicitly contained in the paper by Ashikhmin and
Knill~\cite{ashikhmin01}. 
\begin{proof}
Let $B=\{\beta_1,\dots,\beta_m\}$ denote a basis of
$\F_{q^m}/\F_q$. 
A nondegenerate symmetric form on the $\F_q$-vector space $\F_{q^m}$
is given by $\tr_{q^m/q}(xy)$. It follows that the Gram matrix
$M=(\tr_{q^m/q}(\beta_i\beta_j))_{1\le i,j\le m}$ is nonsingular.
We have $\tr_{q^m/q}(xy)= e_B(x)^tMe_B(y)$ for all $x, y$ in $\F_{q^m}$. 

If $a$ is an element of $\F_{q^m}$, then we denote 
by $e_B(a)$ the coordinate vector in $\F_q^m$ given by 
$ e_B(a) = (a_1,\dots,a_m)$, where $a = \sum_{i=1}^m
a_i\beta_i.$ We define an $\F_p$--vector space isomorphism $\varphi_B$ from
$\F_{q^m}^{2n}$ onto $\F_q^{2nm}$ by 
$$ \varphi_B((a|b)) = ((e_B(a_1),\dots,
e_B(a_n))|(Me_B(b_1),\dots,Me_B(b_n))).$$ It follows from the fact
that $\tr_{q^m/q}(\tr_{q/p}(x))=\tr_{q^m/p}(x)$ holds for all $x$ in
$\F_{q^m}$ and the definition of the isomorphism $\varphi_B$ that
$(a|b)\, \sdual\, (c|d)$ holds in $\F_{q^m}^{2n}$ if and only if
$\varphi_B((a|b)) \,\sdual\, \varphi_B((c|d))$ holds in
$\F_{q^{2nm}}$.

If an $((n,K,d))_{q^m}$ exists, then there exists an additive code
$C\le \F_{q^m}^{2n}$ of size $|C|=q^{nm}/K$ such that $C\le C^\sdual$,
$\swt(C^\sdual\setminus C)=d$ if $K>1$, and $\swt(C^\sdual)=d$ if
$K=1$. Therefore, the code $\varphi_B(C)$ over the alphabet $\F_q$ is
of size $q^{nm}/K$, satisfies $\varphi_B(C)\le \varphi_B(C)^\sdual\le
\F_q^{2nm}$, and $\swt(\varphi_B(C)^\sdual\setminus \varphi_B(C))=d$
if $K>1$ and $\swt(\varphi_B(C)^\sdual)=d$ if $K=1$. Thus, an
$((nm,K,d))_q$ stabilizer code exists. 

The existence of an $((nm,K,d))_q$ stabilizer code implies the
existence of an $((nm,K))_q$ stabilizer code; the claim about the minimum
distance follows from the fact that $\varphi_B^{-1}$ maps each nonzero
block of $m$ symbols to a nonzero symbol in $\F_{q^m}$.
\end{proof}

We notice that if $q$ is even or if $q$ and $m$ are both odd, then
there exists a basis $B$ such that $M$ is the identity matrix; in that
case, $\varphi_B$ simply expands each symbol into coordinates with
respect to~$B$.  If $q$ is odd and $m$ is even, then no such basis
exists.

\section{Conclusions and Open Problems}
We have further developed the theory of nonbinary stabilizer codes.
In the first seven sections, we studied the basic theory of nonbinary
stabilizer codes over finite fields, and introduced Galois-theoretic
methods to clarify the relation between these and more general quantum
codes. In the remaining sections, we derived numerous families of
quantum codes. Table~\ref{tb:families} gives an overview and
summarizes the main parameters of these families.

We should emphasize that it is possible to start with a different
choice of error basis~\cite{knill96a}, and one can develop a similar
theory for such stabilizer codes. For example, one choice leads to
self-orthogonal additive subcodes of $\Z_q^n\times \Z_q^n$ instead of
subcodes of $\F_q^n\times \F_q^n$.  It would be interesting to know
how the stabilizer codes with respect to different error bases
compare.

One central theme in quantum error-correction is the construction of
codes that have a large minimum distance. We were able to show that
the length of an MDS stabilizer code over $\F_q$ cannot exceed
$q^2+1$, except in a few sporadic cases, assuming that the classical
MDS conjecture holds.  An open problem is whether the length $n$ of a
$q$-ary quantum MDS code is bounded by $q^2+1$ for all but finitely
many $n$.

A number of researchers raised the question whether there exist
degenerate quantum codes that can exceed the quantum Hamming bound.
Following Gottesman's lead \cite{gottesman97}, we were able to show
that single and double error-correcting nonbinary stabilizer codes
cannot beat the quantum Hamming bound. We conjecture that no quantum
error-correcting code can exceed the quantum Hamming bound, but a
proof is still elusive.

Finally, we briefly mention some of the topics that we have
deliberately omitted.  We decided not to include tables of the best
known stabilizer codes, but rather make such tables available on the
home page of the second author. We selected code families that are
easily accessible by elementary methods; the interested reader can
find examples of more intricate algebro-geometric constructions
in~\cite{ashikhmin01b,chen01,chen01b,kim04,matsumoto02} and of binary
quantum LDPC codes in~\cite{postol01,mackay04,camara05}.  We did not
include constructive aspects of encoding and decoding circuits, since
encoding circuits are discussed in~\cite{grassl03} and little is known
about the decoding of stabilizer codes. We did not include
combinatorial aspects, but Kim pointed out that there is a forthcoming
book by Glynn, Gulliver, Maks, and Gupta that explores the relation
between binary stabilizer codes and finite geometry.

\paragraph{Acknowledgments.} 
We received numerous comments and suggestions during the preparation
of this manuscript that are much appreciated. In particular, many
thanks to Markus Grassl and Martin R\"otteler for sending us
corrections and suggestions, to Daniel Gottesman, Jon-Lark Kim and
Simon Litsyn for sending us helpful comments and references, and to
Raymond Laflamme and Peter Shor for providing us with historical
background.  We are grateful to Neil Sloane for very fruitful
discussions on MDS codes, and to Gordon Chen, Phil Hemmer, and Suhail
Zubairy for helpful discussions in our quantum computing seminar.

This work would not have been possible without the support of NSF
grant CCR-0218582, NSF CAREER award CCF-0347310, a TITF grant, and a
TEES Select Young Faculty Award.

\begin{landscape}
\begin{table}[htb]
\small 
\begin{center}
\begin{tabular}{@{}l@{}||c|@{}c@{}|@{}c@{}}
Family & $[[n,k,d]]_q$ & Purity & Parameter Ranges and References \\
\hline\hline 
Short MDS & $[[n,n-2d+2,d]]_q$ & pure &$2\leq d\leq \lceil n/2\rceil$, $q^2-1\ge \binom{n}{d}$\\
\hline
Hermitian Hamming & $[[n,n-2m,3]]_q$& pure & $m\ge 2$, $\gcd(m,q^2-1)=1$, $n=(q^{2m}-1)/(q^2-1)$\\
\hline
 Euclidean Hamming & $[[n,n-2m,3]]_q$& pure & $m\ge 2$, $\gcd(m,q-1)=1$, $n=(q^m-1)/(q-1)$\\
\hline
Quadratic Residue  I &$[[n,1,d]]_q$ & pure & $n$ prime, $n\equiv 3 \mod 4$,
$q\not\equiv 0 \mod n$\\
& & & $q$ is a quadratic residue modulo~$n$,
$d^2-d+1\ge n$\\
\hline
Quadratic Residue II &$[[n,1,d]]_q$& pure&
$n$ prime, $n\equiv 1 \mod 4$, $q\not\equiv 0\mod n$\\
& & & $q$ is a quadratic residue modulo~$n$,  $d\ge \sqrt{n}$\\
\hline
Melas  &$[[n,n-4m,\ge
3]]_q$ &pure & $q$ even, $n=q^{2m}-1$, Pure to 3\\
\hline
Euclidean BCH & $[[n,n-2m\lceil(\delta-1)(1-1/q)\rceil,\geq \delta]]_q$&
 pure& $2\leq \delta \leq q^{\lceil m/2 \rceil}-1-(q-2)[m \textrm{ odd}]$\\
& & to $\delta$ & $n=q^m-1$ and $m\geq2$\\
Punctured BCH & $[[d^*(\mu),\geq d^*(\mu)-2m\lceil(\delta-1)(1-1/q)\rfloor,\geq \delta]]_q$ & pure?& $\delta <q^{\lfloor m/2\rceil}-1$, See Corollary~\ref{th:bchpunc}\\
\hline
Hermitian BCH & $[[n,n-2m\lceil(\delta-1)(1-1/q^2)\rceil,\geq \delta]]_q$ & pure&
 $2\leq \delta \leq q^{m}-1$, $n=q^{2m}-1$, Pure to $\delta$ \\
Extended BCH & $[[n+1,n-2m\lceil(\delta-1)(1-1/q^2)\rceil-1,\geq \delta+1]]_q$ & pure&Pure to $\delta+1$\\
\hline Trivial MDS & $ [[n,n-2,2]]_q$ &pure & $n\equiv 0 \mod p$\\
& $[[n,n,1]]_q$& pure& $n\geq 1$\\
\hline Character & $[[n,k(r_2)-k(r_1),\min \{ 2^{m-r_2},2^{r_1+1} \}]]_q$& pure&$n=2^m$,
$q$ odd, $0\leq r_1<r_2\leq m$, $k(r)=\sum_{j=0}^r\binom{m}{j}$\\
\hline CSS GRM &
$[[q^{m},k(\nu_2)-k(\nu_1),\min\{d(\nu_2),d(\nu_1^\perp)\} ]]_q$ &
pure& $k(\nu)= \sum_{j=0}^m (-1)^j
\binom{m}{j}\binom{m+\nu-jq}{\nu-jq}$, $\nu^\perp=m(q-1)-\nu-1$  \\
&$0\leq \nu_1\leq \nu_2\leq m(q-1)-1$ & & $\nu^\perp+1=(q-1)Q+R$, $d(\nu)=(R+1)q^Q$\\
Punctured GRM & $[[d(\mu),\geq k(\nu_2)-k(\nu_1)-(n-d(\mu)),\geq d ]]_q$ &
pure?& $d \geq \min
\{d(\nu_2),d(\nu_1^\perp)$, $0\leq \mu\leq \nu_2-\nu_1$; \cite{klappenecker05p1}\\
\hline Hermitian GRM & $[[q^{2m},q^{2m}-2k(\nu),d(\nu^\perp) ]]_q$& pure &
$ k(\nu) =\sum_{j=0}^m (-1)^j \binom{m}{j}\binom{m+\nu-jq^2}{\nu-jq^2}$, $\nu^\perp=m(q^2-1)-\nu-1$\\ 
&  $0 \leq \nu \leq m(q-1)-1$ & & $\nu^\perp+1 =(q^2-1)Q+R$, $d(\nu)=(R+1)q^{2Q}$\\
Punctured GRM & $[[d(\mu^\perp),\geq d(\mu^\perp)-2k(\nu),\geq d(\nu^\perp) ]]_q$& pure?&
$(\nu+1)q\leq \mu\leq m(q^2-1)-1$; \cite{klappenecker05p1}\\
\hline Punctured MDS & $[[q^2-q\alpha, q^2-q\alpha-2\nu-2,\nu+2]]_q$ &pure & $0\leq \nu\leq q-2$, $0\leq \alpha \leq q-\nu-1$; \cite{klappenecker05p1} \\
\hline Euclidean MDS & $[[n,n-2d+2,d]]_q$& pure&$3\leq n\leq q, 1\leq d\leq
n/2+1$; \cite{grassl03}\\
\hline Hermitian MDS & $[[q^2-s,q^2-s-2d+2,d]]_q$& pure&$1\leq d\leq
q, s=0,1$; \cite{grassl03}\\
\hline  Twisted & $[[q^2+1,q^2-3,3]]_q$& pure?&  \cite{bierbrauer00}\\
\hline
Extended Twisted & $[[q^r,q^r-r-2,3]]_q$ &pure& $r\geq 2$; \cite{bierbrauer00}\\
&$[[n,n-r-2,3]]_q$& pure &
$n=(q^{r+2}-q^3)/(q^2-1)$, $r\geq 1$, $r$ odd; \cite{bierbrauer00}\\
\hline Perfect & $[[n,n-r-2,3]]_q$&pure&
$n=(q^{r+2}-1)/(q^2-1)$, $r\geq 2$, $r$ even; \cite{bierbrauer00}
\end{tabular}
\end{center}
\caption{A compilation of known families of quantum
codes}\label{tb:families}
\end{table}
\end{landscape}

\scriptsize
\def\cprime{$'$}

\end{document}